\documentclass{article}
\usepackage{amssymb}
\usepackage{amsmath}

\newtheorem{theorem}{Theorem}

\newtheorem{proposition}[theorem]{Proposition}

\input{tcilatex}
\begin{document}

\title{{\Large \textbf{Symmetries of spacetimes embedded with an
Electromagnetic String Fluid}}}
\author{Michael Tsamparlis\thanks{%
Email: mtsampa@phys.uoa.gr} \\
%EndAName
{\ \ \textit{Faculty of Physics, Department of
Astronomy-Astrophysics-Mechanics,}}\\
{\ \textit{University of Athens, Panepistemiopolis, Athens 157 83, Greece.}}
\and Antonios Mitsopoulos\thanks{%
Email: antmits@phys.uoa.gr} \\
%EndAName
{\ \textit{Faculty of Physics, Department of
Astronomy-Astrophysics-Mechanics,}}\\
{\ \textit{University of Athens, Panepistemiopolis, Athens 157 83, Greece.}}
\and Andronikos Paliathanasis\thanks{%
Email: anpaliat@phys.uoa.gr} \\
%EndAName
{\ \textit{Instituto de Ciencias F\'{\i}sicas y Matem\'{a}ticas, }}\\
{\textit{Universidad Austral de Chile, Valdivia, Chile.}}\\
{\textit{Institute of Systems Science, Durban University of Technology}}\\
{\textit{Durban 4000, Republic of South Africa.}}}
\date{}
\maketitle

\begin{abstract}
The electromagnetic string fluid (EMSF) is an anisotropic charged string
fluid interacting with a strong magnetic field. In this fluid we consider
the double congruence defined by the 4-velocity of the fluid $u^a$ and the
unit vector $n^a$ along the magnetic field. Using the standard 1+3
decomposition defined by the vector $u^a$ and the 1+1+2 decomposition
defined by the double congruence ${u^a,n^a}$ we determine the kinematic and
the dynamic quantities of an EM string fluid in both decompositions. In
order to solve the resulting field equations we consider simplifying
assumptions in the form of collineations. We decompose the generic quantity $%
L_{X}g_{ab}$ in a trace $\psi$ and and a traceless part $H_{ab}$. Because
all collineations are expressible in terms of the quantity $L_{X}g_{ab}$ it
is possible to compute the Lie derivative of all tensors defined by the
metric i.e. the Ricci tensor, the Weyl tensor etc. This makes possible the
effects of any assumed collineation on the gravitational field equations.
This is done as follows. Using relevant identities of Differential Geometry
we express the quantity $L_{X}R_{ab}$ where $R_{ab}$ is the Ricci tensor in
terms of the two irreducible parts $\psi,H_{ab}$. Subsequently using the
gravitational field equations we compute the same quantity $L_{X}R_{ab}$ in
terms of the Lie derivative of the dynamic variables. We equate the two
results and find the field equations in the form $L_{X}{\text{Dynamic
variable}}=F(\psi,H_{ab},{\text{ dynamic variables}})$. This result is
general and holds for all gravitational systems and in particular for the
EMSF. Subsequently we specialize our study at two levels. We consider the
case of a Conformal Killing Vector (CKV) parallel to $u^a$ and a CKV
parallel to $n^a$. Finally we solve the resulting field equations in the
first case to the Friedman Robertson Walker (FRW) spacetime and in the
second case for the Bianchi I spacetime. In the latter case we find a new
solution of the gravitational field equations.
\end{abstract}

\section{Introduction}

Relativistic magnetohydrodynamics is the main theory which can describes
various phenomena in modern astrophysics, for some recent reviews we refer
the reader in \cite{mhd0,mhd1,mhd2,mhd3}. When a charged plasma enters a
strong magnetic field it is possible the pressures along the magnetic field
and perpendicular to the magnetic field to be unequal. This results in a
physical system which we call an anisotropic electromagnetic string fluid
(EMSF).

In the present work we study the dynamics of an isolated anisotropic
gravitating fluid which for the observers $u^{a}$ ($u_{a}u^{a}=-1)$ has the
energy momentum tensor of the form 
\begin{equation}
T_{ab}=\mu u_{a}u_{b}+p_{\parallel }n_{a}n_{b}+p_{\perp }p_{ab}
\label{Eqn2.1}
\end{equation}%
where

- $n^{a}$ is a unit spacelike vector which is characteristic of the fluid

- $\mu$ is the matter density of the fluid observed by the observers $u^{a} $

- $p_{ab}$ is the tensor projecting normal to both the vectors $u^{a}$ and $%
n^{a}$ defined by the relation:%
\begin{equation}
p_{ab}=h_{ab}-n_{a}n_{b}  \label{NST.2}
\end{equation}%
where 
\begin{equation}
h_{ab}=g_{ab}+u_{a}u_{b}  \label{NST.3}
\end{equation}%
is the tensor projecting normal to the vector $u^{a}.$

This type of fluid is a special case of a string fluid. A general string
fluid is defined \cite{let1979,ss001} as a fluid consisting of the mixture
of a general fluid with energy momentum tensor $T_{ab}$ and a second fluid
which is characterized by an antisymmetric tensor field $F_{ab}$ with energy
momentum tensor $_{F}T_{ab}$ of the form 
\begin{equation}
_{F}T_{ab}=\sigma n_{a}n_{b}  \label{NST.3a}
\end{equation}%
where $\sigma $ \ is a function and the spacelike vector $n^{a}=F^{ab}u_{b}$%
. For certain types of $T_{ab}$ it is possible that the energy momentum
tensor $_{F}T_{ab}+T_{ab}$ of the string fluid has the form (\ref{Eqn2.1}).

There have been many types of sting fluids considered in the literature
involving different combinations of the two fluids (see \cite{Letelier 1980},%
\cite{Letelier 1981},\cite{Letelier 1983}, \cite{MHD_1},\cite{MHD_2},\cite%
{MHD_3}). The recent works on the topic study the string fluid mainly from
the point of view of thermodynamics by considering conserved currents and
the corresponding chemical potentials including - in most of the studies -
the entropy and the temperature. They seem go make no extensive use of the
gravitational field equations and concentrate rather on the catastatic
equations in order to complete the set of the field equations. The problem
with this approach is that the geometry of spacetime does not enter
explicitly into the study therefore one cannot use the results for the
various important spacetime considered in General Relativity and
particularly on the various cosmological models.

Our approach follows the line of older research on string fluids where
emphasis was given in the macroscopic study of a gravitating string fluid
using geometric methods and especially collineations to simplify the
dynamical equations. By a collineation or geometric symmetry of a geometric
object $A$ defined by the metric (not necessarily a tensor) we mean an
equation of the form $L_{X}A=B$ \ where $B$ \ is a tensor field with the
same number and the same symmetries of indices as $A.$ $X$ is the vector
generating the collineation. It can be shown (see \cite{tsbianchi}) that the
quantity $L_{X}A$ can be expressed in terms of the fundamental quantity $%
L_{X}g_{ab}.$ Therefore it is possible to characterize all collineations in
terms of the quantity $L_{X}g_{ab}.$ Furthermore we may consider the
decomposition of the quantity $L_{X}g_{ab}$ in a trace and a traceless part
as follows%
\begin{equation}
L_{X}g_{ab}=2\psi g_{ab}+H_{ab}  \label{NST.4a}
\end{equation}%
where $H_{ab}$ is traceless and characterizes each collineation in terms of
the tensors $\psi, H_{ab}.$

Of particular interest is the quantity $L_{X}R_{ab}$ which enters the
gravitational field equations because allows one to express these equations
in terms of the quantity $L_{X}g_{ab}$ (or the tensors $\psi ,H_{ab})$ so
that for each particular collineation one simply replaces the appropriate
expression of $L_{X}g_{ab}$ and obtains the gravitational equations in a
form that already incorporates the collineation.

Concerning the study of a gravitating string fluid using collineations this
must be based on a scenario which shall be independent of the particular
type of the collineation. In addition it should be systematic in the sense
that it will describe the steps one has to follow in order to get to the
required answers in the easiest and safe way. A first attempt to develop
such a scenario was done in \cite{Tsamparlis M GRG 2006}, however we feel
that it must be stated again briefly in a more systematic way. The proposed
scenario we apply in this work is general and has as follows.

The study of a string fluid requires two main vector fields, the
four-velocity $u^{a}$ and the vector filed $n^{a}$ $(u^{a}n_{a}=0)$
describing a dynamic variable of the fluid. These vector fields define a set
variables which are classified in two sets

a. Kinematic variables which are due to the vector field $u^{a}$

b. Physical variables which are defined by the vector field $n^{a}.$

The kinematic and the dynamical variables are not independent because they
are constrained by the following conditions

i. Certain geometric identities which the vector fields $u^{a},n^{a}$ must
satisfy (e.g. Ricci identity etc)

ii. The gravitational field equations $G_{ab}=T_{ab}+_{F}T_{ab}$ from which
results the conservation law $\left( T^{ab}+_{F}T^{ab}\right) _{;b}=0$

ii. The field equations of the tensor field $F_{ab}$ together with the
constraint $F_{;b}^{ab}=0.$

The standard way to define the dynamic variables of all fluids is the
covariant decomposition of the tensor fields defining the fluid by means of
the 1+3 decomposition defined by the projection tensor $h_{ab}$ in (\ref%
{NST.3}) and the 1+1+2 decomposition defined by the screen projection tensor 
$p_{ab}$ in (\ref{NST.2}).

Subsequently all above conditions a.-b. and i. - iii. must be expressed 1+3
and 1+1+2 decomposed in order to define the physical variables and
subsequently to obtain a set of equations which incorporate fully the
dynamics of the string fluid in all spacetimes and for all possible
collineations. As expected\ the set of the final equations is not enough for
the determination of all dynamical variables and one has to introduce new
assumptions which are the equations of state and possibly other additional
physical assumptions.

We demonstrate the above scenario to a particular string fluid which has
been considered in various forms in the literature, see \cite{let1979,ss001}
and references therein. This string fluid consists of a mixture of a charged
perfect fluid interacting with an electromagnetic field which is described
in the RMHD approximation with vanishing electric field and infinite
conductivity. This particular string fluid in the following we shall call
the electromagnetic string fluid (EMSF). Physically an EMSF occurs for
example when a charged plasma enters a strong magnetic field where it is
possible the pressures along the magnetic field and perpendicular to the
magnetic field are unequal. In this case the EM\ field contains a magnetic
field only $H^{a}$ and $F_{ab}=\eta _{abcd}H^{c}u^{d}$ whereas the vector $%
n^{a}=H^{a}/H.$

Our purpose is to provide a useful tool for the people who would like to
work on this direction in the future. Inevitably in this approach we shall
have to recall pieces of results from previous works which have required
long and tedious calculations which there is no point to be repeated (see 
\cite{Tsamparlis Mason 1990}).

The structure of the paper is as follows. In Section \ref{section2} for the
convenience of the reader we review briefly the definition of physical
variables of the string fluid using the the 1+3 and 1+1+2 decompositions.
The EMSF is defined in Section \ref{section3} where we discuss the 1+1+2
decomposition of Maxwell equations. Furthermore, the gravitational field
equations for the model are presented in Section \ref{section4}. In Sections %
\ref{section5} and \ref{section6} we find the kinematic and the dynamic
conditions for the gravitational field equations when the spacetime admits a
timelike or a spacelike CKV. We demonstrate our results in Section \ref%
{section7} where we apply the constraint conditions in the case of Bianchi I
spacetime. Finally, in Section \ref{section8} we draw our conclusions.

\section{The definition of the physical variables}

\label{section2}

The definition of the dynamical variables of a relativistic fluid is done
with the use of the 1+3 and the 1+1+2 decomposition of the characteristic
fields $u^{a},n^{a}$ of the fluid$.$ The 1+3 decomposition generated by $%
u^{a}$ can be found among others in the early (and excellent) paper of Ellis 
\cite{ellis2} whereas the 1+1+2 in \cite{Saridakis Tsamparlis 1991}, \cite%
{Tsamparlis Mason 1990}{\Large \ kai alla.} In the following we review
briefly the application of these decompositions for the case of a string
fluid.

\subsection{The 1+3 decomposition}

Consider a spacetime with metric $g_{ab}$ and a fluid of observers with
four-velocity $u^{a}$ ($u^{a}u_{a}=-1$). $u^{a}$ defines the projection
operator $h_{ab}=g_{ab}+u_{a}u_{b}$ with respect to which all geometric
objects defined on $M$ can be $1+3$ decomposed.

The kinematical variables are defined by the the 1+3 decomposition of 
\begin{equation*}
u_{a;b}=-\dot{u}_{a}u_{b}+\omega _{ab}+\underbrace{\sigma _{ab}+\frac{1}{3}%
\theta h_{ab}}_{=\theta _{ab}}
\end{equation*}
$\omega _{ab}=h_{a}^{c}h_{b}^{d}u_{[c;d]}$ is called the vorticity tensor, $%
\theta _{ab}=h_{a}^{c}h_{b}^{d}u_{(c;d)}$, $\theta =\theta
_{a}^{a}=h^{ab}u_{a;b}=u^{a}{}_{;a}$ is called the expansion (isotropic
strain) and $\sigma _{ab}=\theta _{ab}-\frac{1}{3}\theta h_{ab}=\left[
h_{(a}^{c}h_{b)}^{d}-\frac{1}{3}h^{cd}h_{ab}\right] u_{c;d}$ is called the
shear stress tensor or simply the shear. From the vorticity tensor one
defines the vorticity vector 
\begin{equation*}
\omega ^{a}=\frac{1}{2}\eta ^{abcd}u_{b;c}u_{d}.
\end{equation*}%
At the level of dynamics the 1+3 decomposition concerns the energy momentum
tensor $T_{ab}$ of a fluid in $M$ and defines the physical quantities of the
fluid as observed by the observers $u^{a}$ as follows 
\begin{equation}
T_{ab}=\mu u_{a}u_{b}+ph_{ab}+2q_{(a}u_{b)}+\pi _{ab}.  \label{EqnSS.0}
\end{equation}%
The variables $\mu ,p,q^{a},\pi _{ab}$ have the following physical
interpretation:

a. The scalars $\mu = T_{ab} u^a u^b$ and $p = \frac{1}{3} h^{ab} T_{ab}$
correspond respectively to the energy (mass) density and the isotropic
pressure of the fluid.

b. The spacelike vector $q_a = - h^d_a T_{dc} u^c$ is the energy (heat) flux
in the three space defined by the projection tensor $h_{ab}$.

c. $\pi_{ab} = \left( h^c_a h^d_b - \frac{1}{3} h^{cd} h_{ab} \right) T_{cd}$
is the traceless $(\pi _{a}^{a}=0)$ stress tensor (measures the anisotropy).

The fluids are classified according to the dynamical variables in dust $%
(p=q_{a}=\pi _{ab}=0),$ perfect fluid $(q_{a}=\pi _{ab}=0),$ heat conducting
fluid $(q_{a}\neq 0)$ and anisotropic fluid $(\pi _{ab}\neq 0).$

At the same level one considers the 1+3 decomposition of the conservation
equation $T_{\text{ \ \ };b}^{ab}=0$ which leads to the following two
equations ( see for example \cite{Saridakis Tsamparlis 1991})

\begin{eqnarray}
\overset{.}{\mu }+(\mu +p)\theta +\pi ^{ab}\sigma _{ab}+q_{;a}^{a}+q^{a}%
\overset{.}{u}_{a} &=&0  \label{CE.1} \\
(\mu +p)\overset{.}{u}_{a} + h^c_a (p_{,c} + \pi^b_{c;b} + \overset{.}{q}%
_{c}) + \left( \omega_{ac} + \sigma_{ac} + \frac{4}{3} \theta h_{ac} \right)
q^c &=& 0.  \label{CE.2}
\end{eqnarray}

Finally the 1+3 decomposition of the Ricci identity $%
u_{a;bc}-u_{a;cb}=R^{d}{}_{abc}u_{d}$ leads to two more sets of equations
which are called the propagation and the constraint equations which are
explicitly stated in \cite{ellis2}.

\subsection{The 1+1+2 decomposition}

The pair of vectors $u^{a},n^{a}$ ($n^{a}n_{a}=1,u^{a}n_{a}=0$) constitutes
a double congruence and defines the new projection operator 
\begin{equation*}
p_{ab}=h_{ab}-n_{a}n_{b}
\end{equation*}%
which projects normal to both vectors $u^{a},n^{a}$ on the 2-dimensional
space called the screen space.

The 1+1+2 decomposition defines the \textquotedblleft
kinematic\textquotedblright\ variables of the spacelike vector field $n^{a}$
(see \cite{Mason Tsamparlis 1985}, \cite{Saridakis Tsamparlis 1991} and
references cited therein): 
\begin{equation}
n_{a;b}=A_{ab}+\overset{\ast }{n}_{a}n_{b}-\dot{n}_{a}u_{b}+u_{a}\left[
n^{c}u_{c;b}+(n^{c}\dot{u}_{c})u_{b}-(n^{c}\overset{\ast }{u}_{c})n_{b}%
\right]  \label{N.17}
\end{equation}%
where $\dot{s}\equiv s_{...;a}u^{a}$, $\overset{\ast }{s}\equiv
s_{...;a}n^{a}$ and $A_{ab}=p_{a}^{c}p_{b}^{d}n_{c;d}$. We decompose the
screen tensor $A_{ab}$ into its irreducible parts (the kinematic variables
of $n^{a}$): 
\begin{equation}
A_{ab}=\mathcal{S}_{ab}+\mathcal{R}_{ab}+\frac{1}{2}\mathcal{E}p_{ab}
\label{N.17a}
\end{equation}%
where $\mathcal{S}_{ab}=\mathcal{S}_{ba}$, $\mathcal{S}_{\text{\ }b}^{b}=0$
is the traceless part (screen shear), $\mathcal{R}_{ab}=-\mathcal{R}_{ba}$
is the antisymmetric part (screen rotation) and $\mathcal{E}$ is the trace
(screen expansion). We have the defining relations: 
\begin{align}
\mathcal{S}_{ab}& =\left( p_{a}^{c}p_{b}^{d}-\frac{1}{2}p^{cd}p_{ab}\right)
n_{(c;d)}  \label{N.18} \\
\mathcal{R}_{ab}& =p_{a}^{c}p_{b}^{d}n_{[c;d]}  \label{N.19} \\
\mathcal{E}& \mathcal{=}p^{cd}n_{c;d}=n^{c}{}_{;c}+\dot{n}^{c}u_{c}.
\label{N.20}
\end{align}%
One can define also the screen rotation vector $\mathcal{R}^{a}=\frac{1}{2}%
\eta ^{abcd}\mathcal{R}_{bc}u_{d}$.

The $u^{a}-$term in (\ref{N.17}) can be written: 
\begin{equation}
-N_{b}+2\omega _{cb}n^{c}+p_{b}^{c}\overset{.}{n}_{c}  \label{N.21}
\end{equation}%
where 
\begin{equation}
N_{b}=p_{bc}\left( \dot{n}^{c}-\overset{\ast }{u^{c}}\right) =p_{bc}L_{%
\mathbf{u}}n^{c}  \label{N.22}
\end{equation}%
is called the Greenberg vector. This vector is important because it vanishes
iff the vector fields $u^{a},n^{a}$ are surface forming (that is iff $L_{%
\mathbf{u}}n^{b}=Au^{b}+Bn^{b})$. From the kinematics point of view the
vector $N_{a}$ vanishes iff the vector field $n^{a}$ is \textquotedblleft
frozen\textquotedblright\ along the observers $u^{a}.$

Using the Greenberg vector, identity (\ref{N.17}) \ is written:%
\begin{equation*}
n_{a;b}=A_{ab}+\overset{\ast }{n}_{a}n_{b}-\overset{.}{n}%
_{a}u_{b}+u_{a}p_{b}^{c}\left( \overset{.}{n}_{c}+2\omega
_{dc}n^{d}-N_{c}\right).
\end{equation*}

Concerning the physical variables the energy momentum tensor is 1+1+2
decomposed as follows 
\begin{align*}
T_{ab}& =\mu u_{a}u_{b}+2\kappa u_{(a}n_{b)}+2Q_{(a}u_{b)}+\gamma
n_{a}n_{b}+2P_{(a}n_{b)}+\frac{1}{2}\alpha p_{ab}+D_{ab} \\
& =\mu u_{a}u_{b}+ph_{ab}+2\kappa u_{(a}n_{b)}+2Q_{(a}u_{b)}+\bar{\gamma}%
\left( n_{a}n_{b}-\frac{1}{2}p_{ab}\right) +2P_{(a}n_{b)}+D_{ab}
\end{align*}%
which introduces the new dynamical variables $\kappa ,\gamma ,\alpha ,\bar{%
\gamma},Q_{a},P_{a},D_{ab}$ which correspond to the physical variables
observed by the observers $u^{a}$. They are given by the formulae 
\begin{align}
\kappa & =-T_{ab}u^{a}n^{b}  \label{N.12} \\
\gamma & =T_{ab}n^{a}n^{b},\quad \bar{\gamma}=\pi _{ab}n^{a}n^{b}=\gamma -p
\label{N.13} \\
\alpha & =p^{ab}T_{ab}=3p-\gamma  \label{N.14} \\
Q_{a}& =-T_{bc}u^{b}p_{a}^{c},\quad P_{a}=T_{bc}n^{b}p_{a}^{c},\quad
q_{a}=\kappa n_{a}+Q_{a}  \label{N.15} \\
D_{ab}& =\left( p_{a}^{c}p_{b}^{d}-\frac{1}{2}p^{cd}p_{ab}\right) T_{cd}.
\label{N.16}
\end{align}

The physical meaning of each of the new physical variables is the following

a. The scalar $\kappa$ and the screen vector $Q_{a}$ are related to the heat
conduction of the fluid

b. The scalar $\gamma$, the screen vector $P_{a}$, and the traceless screen
tensor $D_{ab}$ have to do with the anisotropy of the fluid.

The 1+1+2 dynamical variables are related to the 1+3 dynamical variables as
follows 
\begin{align}
q^{a} & = \kappa n^{a} + Q^{a}  \label{N.10} \\
\pi _{ab} & = \bar{\gamma} \left( n_{a} n_{b} - \frac{1}{2} p_{ab} \right) +
2 P_{(a} n_{b)} + D_{ab}.  \label{N.11}
\end{align}

The above decompositions are general and hold for all fluids.

\subsection{The string fluid defined by the electromagnetic field}

The energy momentum tensor (\ref{Eqn2.1}) can be rewritten as:

\begin{equation}
T_{ab}=\mu u_{a}u_{b}+\frac{1}{3}(p_{\parallel }+2p_{\perp
})h_{ab}+(p_{\perp }-p_{\parallel })\left( \frac{1}{3}h_{ab}-n_{a}n_{b}%
\right) .  \label{EqnSS.2}
\end{equation}%
from which follows%
\begin{eqnarray}
p &=&\frac{1}{3}(p_{\parallel }+2p_{\perp }),  \label{EqnSS.3} \\
\pi _{ab} &=&(p_{\perp }-p_{\parallel })\left( \frac{1}{3}%
h_{ab}-n_{a}n_{b}\right) ,  \label{EqnSS.4} \\
q_{a} &=&0.  \label{EqnSS.4a}
\end{eqnarray}

It follows from (\ref{EqnSS.2}) that (\ref{Eqn2.1}) corresponds to an
anisotropic fluid with vanishing heat flux. Furthermore we note that $n^{a}$
is an eigenvector of the anisotropic stress tensor $\pi _{ab}$ with
eigenvalue $-\frac{2}{3} (p_{\perp }-p_{\parallel}).$ We assume $p_{\perp
}-p_{\parallel }\neq 0$ otherwise the string fluid reduces to a perfect
fluid with energy momentum tensor $T_{ab}=p_{\perp} g_{ab}$ which has the
unphysical equation of state $\mu +p_{\perp}=0$.

\bigskip Furthermore, in the 1+1+2 decomposition the tensor $\pi_{ab}$ is
written as

\begin{equation}
\pi _{ab}=-\frac{2}{3}(p_{\perp }-p_{\parallel })\left( n_{a}n_{b}-\frac{1}{2%
}p_{ab}\right) ,  \label{EqnSS.5}
\end{equation}%
from which follows that the only nonvanishing irreducible part of $\pi _{ab}$
is:%
\begin{equation}
\bar{\gamma}=-\frac{2}{3}(p_{\perp }-p_{\parallel }).
\end{equation}%
Hence, we conclude that the string fluid defined by (\ref{EqnSS.2}) \ is the
\textquotedblleft simplest\textquotedblright\ anisotropic fluid.

An important example of a string fluid is the electromagnetic field in the
RMHD approximation with infinite conductivity and vanishing electric
field(see \cite{MHD_1},\cite{MHD_2},\cite{MHD_3}).

Indeed in this approximation the electromagnetic tensor $F_{ab}$ is given by
the expression 
\begin{equation}
F_{ab}=\eta _{abcd}u^{c}H^{d}  \label{EqnSS.5b}
\end{equation}%
where $H^{a}$ is the magnetic field and the vector $n^{a}=H^{a}/H$ is the
unit vector in the direction of the magnetic field.

The Minkowski energy momentum tensor of the electromagnetic field $%
T_{EM}^{ab}$ is given by 
\begin{equation}
_{EM}T^{ab}=\lambda \left( F^{ac}F_{\text{ \ }c}^{b}-\frac{1}{4}%
g^{ab}F_{cd}F^{cd}\right)  \label{ME.14}
\end{equation}%
where $\lambda $ is a constant. Using Maxwell equations one shows that 
\begin{equation}
_{EM}T_{\text{ \ \ \ };b}^{ab}=-F^{ab}J_{b}.  \label{ME.15}
\end{equation}%
Replacing in (\ref{ME.14}) $F_{ab}$ from (\ref{EqnSS.5b}) we find%
\begin{equation}
_{EM}T_{ab}=\frac{1}{2}\lambda H^{2}u_{a}u_{b}+\frac{1}{6}\lambda
H^{2}h_{ab}+\lambda H^{2}\left( \frac{1}{3}h_{ab}-n_{a}n_{b}\right) .
\label{EqnSS.6}
\end{equation}

Replacing $\ h_{ab}=p_{ab}+n_{a}n_{b}$ \ we find%
\begin{equation}
_{EM}T_{ab}=\frac{1}{2}\lambda H^{2}(u_{a}u_{b}-n_{a}n_{b})+\frac{1}{2}%
\lambda H^{2}p_{ab}  \label{EqnSS.7}
\end{equation}%
which defines an anisotropic fluid with energy momentum tensor (\ref{Eqn2.1}%
) with $-\mu =p_{\parallel }=-p_{\perp }=-\frac{1}{2}\lambda H^{2}$. An
anisotropic fluid with energy momentum tensor of the form (\ref{Eqn2.1}) \
such that $-\mu =p_{\parallel }$we shall call perfect string fluid. From (%
\ref{EqnSS.3}) - (\ref{EqnSS.4a}) follows that for a perfect fluid 
\begin{eqnarray}
\mu &=&\frac{1}{2}\lambda H^{2},  \label{EqnSS.10} \\
p &=&\frac{1}{6}\lambda H^{2},  \label{EqnSS.11} \\
\pi _{ab} &=&\lambda H^{2}\left( \frac{1}{3}h_{ab}-n_{a}n_{b}\right) ,
\label{EqnSS.12} \\
q^{a} &=&0  \label{EqnSS.13}
\end{eqnarray}

Therefore, the equation of state parameter for the EMSF is always $p=\frac{1%
}{3}\mu $, and while $\mu \geq 0$, then necessarily~$p\geq 0$\ and the EMSF
satisfies the strong energy condition. 

\section{The electromagnetic string fluid (EMSF)}

\label{section3}

We consider the dynamical system consisting of a charged perfect fluid with
isotropic pressure $p$ and energy density $\rho $ which interacts with the
electromagnetic field in the RMHD approximation with infinite conductivity
and vanishing electric field. Physically this situation is considered to be
the case in various plasmas \cite{plasma}.

Due to the interaction of the fluid with the electromagnetic filed it is
possible that the magnetic field produces a different fluid pressure
perpendicular and parallel to the magnetic field therefore the perfect fluid
becomes an anisotropic fluid with pressure distribution $p_{\parallel
}n_{a}n_{b}+p_{\perp }p_{ab}.$ The energy momentum tensor of the interacting
fluid is then%
\begin{equation}
T_{ab}=\left( \rho +\frac{1}{2}\lambda H^{2}\right) u_{a}u_{b}+\left(
p_{\parallel }-\frac{1}{2}\lambda H^{2}\right) n_{a}n_{b}+\left( p_{\perp }+%
\frac{1}{2}\lambda H^{2}\right) p_{ab}.  \label{EqnSS.8}
\end{equation}

The interacting fluid is not a perfect string fluid. For this to be the case
the following condition must be satisfied

\begin{equation}
\rho +\frac{1}{2}\lambda H^{2}=-\left( p_{\parallel }-\frac{1}{2}\lambda
H^{2}\right) \implies \rho =-p_{\parallel }.  \label{EqnSS.15}
\end{equation}

With this condition assumed the energy momentum of the EMSF is%
\begin{equation}
T_{ab}=\left( \rho +\frac{1}{2}\lambda H^{2}\right) u_{a}u_{b}-\left( \rho +%
\frac{1}{2}\lambda H^{2}\right) n_{a}n_{b}+\underbrace{\left( p_{\perp }+%
\frac{1}{2}\lambda H^{2}\right) }_{\equiv q}p_{ab}.  \label{EqnSS.15a}
\end{equation}%
Then 1+3 decomposition gives 
\begin{eqnarray}
\mu &=&\rho +\frac{1}{2}\lambda H^{2},  \label{EqnSS.17} \\
p &=&\frac{1}{3}\left( 2p_{\perp }-\rho +\frac{1}{2}\lambda H^{2}\right) ,
\label{EqnSS.18} \\
q^{a} &=&0,  \label{EqnSS.19} \\
\overline{\pi }_{ab} &=&(\rho +p_{\perp }+\lambda H^{2})\left( \frac{1}{3}%
h_{ab}-n_{a}n_{b}\right)  \label{EqnSS.20}
\end{eqnarray}%
Concerning the 1+1+2 dynamical variables we have that for the EMSF 
\begin{equation*}
\mu =\rho +\frac{1}{2}\lambda H^{2},\enskip\kappa =0,\enskip Q_{a}=0,\enskip%
\gamma =-\rho -\frac{1}{2}\lambda H^{2},\enskip P_{a}=0,\enskip\alpha
=2p_{\perp }+\lambda H^{2},\enskip D^{ab}=0.
\end{equation*}%
But 
\begin{equation*}
\alpha =3p-\gamma \implies p=\frac{2}{3}\left( p_{\perp }+\frac{1}{2}\lambda
H^{2}\right) +\frac{\gamma }{3}
\end{equation*}%
therefore 
\begin{equation*}
\bar{\gamma}=\gamma -p\implies \bar{\gamma}=-\frac{2}{3}\left( \rho
+p_{\perp }+\lambda H^{2}\right) .
\end{equation*}%
Finally we note the relation 
\begin{equation*}
\mu +q=\rho +p_{\perp }+\lambda H^{2}
\end{equation*}%
which is useful in the calculations. Here $q=p_{\perp }+\frac{1}{2}\lambda
H^{2}$ and thus for the EMSF the energy momentum tensor is written $%
T_{ab}=\mu (u_{a}u_{b}-n_{a}n_{b})+qp_{ab}$ which is the general form for
any perfect string fluid.

One direction in which the string fluids have been studied is the
simplification of the field equations for various types of collineations of
spacetime \cite{Yavuz Yilmaz (1997)}, \cite{Yilmaz 2001}, \cite{Baysal Camci
et all 2002}, \cite{Baysal Yilmaz 2002}, \cite{U Camci 2002}, \cite{Sharif M
Sheikh U 2005} \cite{Tsamparlis M GRG 2006}. In the next sections we extend
these studies to the case of the EMSF.

\subsection{The Ricci tensor of the EMSF}

We consider Einstein field equations in the form $R_{ab}=$ $T_{ab}+\left(
\Lambda -\frac{1}{2}T\right) g_{ab}$ where $T\equiv T_{a}^{a}$ and compute $%
R_{ab}$ in terms of the string fluid variables: We find:%
\begin{equation}
R_{ab}=\left( p_{\perp }+\frac{1}{2}\lambda H^{2}-\Lambda \right)
(u_{a}u_{b}-n_{a}n_{b})+\left( \rho +\frac{1}{2}\lambda H^{2}+\Lambda
\right) p_{ab}.  \label{EqnSS.25}
\end{equation}%
We note immediately that $R_{ab}$ is found from $T_{ab}$ given by (\ref%
{Eqn2.1}) for $p_{\parallel} = - \mu$ if we interchange $\mu \leftrightarrow
p_{\perp }+\frac{1}{2}\lambda H^{2}-\Lambda ,\;\;p_{\perp }\leftrightarrow
\rho +\frac{1}{2}\lambda H^{2}+\Lambda $ and vice versa. This is a useful
observation because it allows us to compute various results for $%
R_{ab}/T_{ab}$ and write down the answer for the corresponding quantities
for $T_{ab}/R_{ab}$ by interchanging the string variables as indicated
above. For example the 1+3 decomposition of $R_{ab}$ is written directly
from (\ref{EqnSS.2}) as follows: 
\begin{equation}
R_{ab}=\left( p_{\perp }+\frac{1}{2}\lambda H^{2}-\Lambda \right) u_{a}u_{b}+%
\frac{1}{3}\left( 2\rho -p_{\perp }+\frac{1}{2}\lambda H^{2}+3\Lambda
\right) h_{ab}+(\rho +p_{\perp }+\lambda H^{2})\left( \frac{1}{3}%
h_{ab}-n_{a}n_{b}\right) .  \label{EqnSS.26}
\end{equation}

\subsection{The conservation equations for the EMSF in the 1+1+2
decomposition}

In the case of the EMSF the conservation equations (\ref{CE.1}),(\ref{CE.2})
are simplified as follows%
\begin{eqnarray}
\overset{.}{\mu }+(\mu +q)\left( \frac{2}{3}\theta -\sigma
_{ab}n^{a}n^{b}\right) &=&0  \label{CE.3} \\
(\mu +q)\left[ \overset{.}{u}_{a}-(\mathcal{E}-\overset{.}{n}%
_{b}u^{b})n_{a}-h_{a}^{b}\overset{\ast }{n}_{b}\right] +p_{a}^{b}q_{,b}-%
\overset{\ast }{\mu }n_{a} &=&0.  \label{CE.4}
\end{eqnarray}

Furthermore, by projecting the second equation along $n^{a}$ and using of
the tensor $p_{b}^{a},$ we get the two equations%
\begin{eqnarray}
\overset{\ast }{\mu }+(\mu +q)\mathcal{E} &=&0 \\
p_{a}^{b}\left[ q_{,b}+(\mu +q)(\overset{.}{u}_{b}-\overset{\ast }{n}_{b})%
\right] &=&0.
\end{eqnarray}%
Replacing the energy density $\mu$ and the heat coefficient $q$ from the
expressions of the previous section we find that the 1+1+2 decomposition of
the conservation equation for an EMSF are 
\begin{eqnarray}
\overset{.}{\rho }+\lambda H\overset{.}{H}+(\rho +p_{\perp }+\lambda H^{2})
\left( \frac{2}{3}\theta -\sigma _{ab}n^{a}n^{b} \right) &=& 0
\label{EqnSS.37} \\
\overset{\ast }{\rho }+\lambda H\overset{\ast }{H}+(\rho +p_{\perp }+\lambda
H^{2})\mathcal{E} &=&0  \label{EqnSS.38} \\
p_{a}^{b}\left[ p_{\perp }{}_{,b}+\lambda HH_{,b}+(\rho +p_{\perp }+\lambda
H^{2})(\overset{.}{u}_{b}-\overset{\ast }{n}_{b})\right] &=&0.
\label{EqnSS.39}
\end{eqnarray}%
\qquad

These equations are independent of any other assumptions which one might do
concerning the fluid, including the symmetries. Therefore for each
additional assumption (including the symmetry assumptions) the conservation
equations take a different form and in that form supplement the rest of the
field equations as constraint equations.

We continue our analysis with the 1+1+2 decomposition of Maxwell equations.

\subsection{Maxwell equations in the 1+3 and the 1+1+2 formalisms}

Maxwell equations are%
\begin{equation}
F_{[ab;c]}=0,F_{\;\;;b}^{ab}=J^{a}  \label{ME.1}
\end{equation}%
where $F^{ab}$ is the electromagnetic field tensor and $J^{a}$ is the
4-current. The 4-current and the electromagnetic field tensor in the 1+3
decomposition are decomposed as follows 
\begin{eqnarray}
J^{a} &=&eu^{a}+j^{a}  \label{ME.2} \\
F^{ab} &=&u^{a}E^{b}-u^{b}E^{a}+\eta ^{abcd}H_{c}u_{d}  \label{ME.3}
\end{eqnarray}%
where $\eta ^{abcd}$ \ is the alternating tensor\footnote{%
In Minkowski spacetime the alternating tensor is defined as follows:
\par
\begin{equation*}
\eta ^{abcd}=\eta ^{\lbrack abcd]},\eta ^{0123}=(-g)^{-1/2}
\end{equation*}%
\par
where $g=\det (g_{ab}).$It satisfies the properties:
\par
\begin{equation*}
\eta ^{abcd}\eta _{arst}=-3!\delta _{r}^{[b}\delta _{s}^{c}\delta
_{t}^{d]},\eta ^{abcd}\eta _{abst}=-4\delta _{s}^{[c}\delta _{t}^{d]}
\end{equation*}%
\par
In the Euclidian 3-d space the alternating tensor is defined as follows:
\par
\begin{equation*}
\eta ^{\mu \nu \rho }=\eta ^{\lbrack \mu \nu \rho ]},\eta ^{123}=h^{-1/2}
\end{equation*}%
where $h=\det (h_{\mu \nu }).$ It satisfies the properties:
\par
\begin{equation*}
\eta ^{\mu \nu \rho }\eta _{\mu \sigma \tau }=2\delta _{\sigma }^{[\nu
}\delta _{\tau }^{\rho ]},\eta ^{\mu \nu \rho }\eta _{\mu \nu \tau }=2\delta
_{\tau }^{\rho }.
\end{equation*}%
} and the various physical quantities introduced are (a) $e$ the charge
density (b) $j^{a}$ the conduction current (c) $E^{a}$ the electric field
and (d) $H^{a}$ the magnetic field, \ all these quantities measured by the
observer $u^{a}.$ Inverting (\ref{ME.2}), (\ref{ME.3}) we find%
\begin{eqnarray}
e &=&-u^{a}J_{a},j^{a}=h_{b}^{a}J^{b}  \label{ME.4} \\
E^{a} &=&F^{ab}u_{b},H^{a}=\frac{1}{2}\eta ^{abcd}F_{bc}u_{d}.  \label{ME.5}
\end{eqnarray}%
\textbf{At this point we would like to note that from the f{rom the 1+3
decomposition of $F_{ab}$ wrt $u^{a}$ it follows that the magnetic part $%
H^{a}$ is orthogonal to $u^{a}$, i.e. $H^{a}u_{a}=0$. Thereupon, the chosen
direction $n^{a}=H^{a}/H$ satisfies the relation $n^{a}u_{a}=0$. That's why
the direction of the magnetic field wrt the observer $u^{a}$ is always taken
orthogonal.}}

Taking into account the 1+3 kinematic variables Maxwell equations are 1+3
decomposed wrt the observer $u^{a}$ into the following constraint and
propagation equations (see \cite{ellis2}, \cite{SR}) 
\begin{eqnarray}
h_{b}^{a}H_{;a}^{b} &=&2\omega ^{a}E_{a}  \label{ME.6} \\
h_{b}^{a}E_{;a}^{b} &=&e-2\omega ^{a}H_{a}  \label{ME.7} \\
h_{b}^{a}\overset{.}{H}^{b} &=&u_{;b}^{a}H^{b}-\theta H^{a}-I^{a}(E)
\label{ME.8} \\
h_{b}^{a}\overset{.}{E}^{b} &=&u_{;b}^{a}E^{b}-\theta E^{a}+I^{a}(H)-j^{a}
\label{ME.9}
\end{eqnarray}%
where:%
\begin{eqnarray}
I^{a}(E) &=&\eta ^{abcd}u_{b}(\overset{.}{u}_{c}E_{d}-E_{c;d})  \label{ME.10}
\\
I^{a}(H) &=&\eta ^{abcd}u_{b}(\overset{.}{u}_{c}H_{d}-H_{c;d})  \label{ME.11}
\end{eqnarray}%
and $\omega ^{a}=\frac{1}{2}\eta ^{abcd}u_{b;c}u_{d}$ and $\theta
=u_{;a}^{a} $ are the the vorticity vector and the expansion of the fluid as
measured by the observers $u^{a}$ . A dot over a symbol denotes covariant
differentiation wrt $u^{a}$ (i.e. along the fluid particle world line).

If we operate on (\ref{ME.10}) and (\ref{ME.11}) with $\eta ^{abcd}u_{d}$ \
then a direct calculation yields the following two mathematical identities:%
\begin{eqnarray}
E_{[r;s]} &=&u_{[r}\overset{.}{E}_{s]}+\overset{.}{u}%
_{[r}E_{s]}+u^{t}E_{t;[r}u_{s]}+\frac{1}{2}\eta _{rstm}u^{t}I^{m}(E)
\label{ME.12} \\
H_{[r;s]} &=&u_{[r}\overset{.}{H}_{s]}+\overset{.}{u}%
_{[r}H_{s]}+u^{t}H_{t;[r}u_{s]}+\frac{1}{2}\eta _{rstm}u^{t}I^{m}(H).
\label{ME.13}
\end{eqnarray}

From the identity (\ref{ME.13}) \ one computes the screen rotation of the
magnetic field lines. The result is: 
\begin{eqnarray}
\mathcal{R}_{ab} &=& p_{a}^{c}p_{b}^{d}n_{[c;d]} = \frac{1}{H}%
p_{a}^{c}p_{b}^{d}H_{[c;d]}  \notag \\
&=&-\frac{1}{2H}p_{a}^{c}p_{b}^{d}\eta _{cdrs}I^{r}(H)u^{s}.  \label{ME.23}
\end{eqnarray}

\begin{proposition}
\label{Spacelike MHD Rotation Vector}The screen rotation vector of the
magnetic field lines is proportional to the magnetic field as follows:%
\begin{equation}
\mathcal{R}^{a}=-\frac{H_{c}I^{c}(H)}{2H^{3}}H^{a}  \label{ME.24}
\end{equation}
\end{proposition}

Proof

Expanding $p_{a}^{c},$ $p_{b}^{d}$ in (\ref{ME.23}) \ we get:%
\begin{equation*}
\mathcal{R}_{ab}=-\frac{1}{2H}\eta _{abrs}I^{r}(H)u^{s}-\frac{1}{H^{3}}%
H_{[a}\eta _{b]crs}H^{c}I^{r}(H)u^{s}
\end{equation*}

We operate with $\eta ^{abpq}$ on both sides and find:%
\begin{eqnarray*}
-\frac{1}{2H}\eta ^{abpq}\eta _{abrs}I^{r}(H)u^{s} &=&\frac{1}{H}%
[I^{p}(H)u^{q}-I^{q}(H)u^{p}] \\
-\frac{1}{H^{3}}\eta ^{abpq}H_{[a}\eta _{b]crs}H^{c}I^{r}(H)u^{s} &=&\frac{1%
}{H}\left[ -I^{p}(H)u^{q}+I^{q}(H)u^{p}\right] + \frac{H_{c}I^{c}(H)}{H^{3}}%
\left[ H^{p}u^{q}-H^{q}u^{p}\right] .
\end{eqnarray*}

Thus 
\begin{equation}
\eta ^{abpq}\mathcal{R}_{ab}=\frac{H_{c}I^{c}(H)}{H^{3}}\left[
H^{p}u^{q}-H^{q}u^{p}\right] .  \label{ME.25}
\end{equation}

In terms of the screen rotation vector $\mathcal{R}^{a}=\frac{1}{2}\eta
^{abcd}u_{b}\mathcal{R}_{cd}$ equation (\ref{ME.25}) is written:%
\begin{equation*}
\mathcal{R}^{a}=-\frac{H_{c}I^{c}(H)}{2H^{3}}H^{a}
\end{equation*}

which completes the proof.

From Proposition \ref{Spacelike MHD Rotation Vector} we infer that the
screen rotation of the magnetic field congruence vanishes iff $H_{c}I^{c}(H)$
$=0$.

\subsection{Maxwell equations in the RMHD approximation}

In the RMHD approximation with infinite electric conductivity and vanishing
electric field Maxwell equations become:

\begin{eqnarray}
h_{b}^{a}H_{;a}^{b} &=&0  \label{ME.16} \\
e&=&2\omega ^{a}H_{a}  \label{ME.17} \\
h_{b}^{a}\overset{.}{H}^{b} &=&u_{;b}^{a}H^{b}-\theta H^{a}  \label{ME.18} \\
I^{a}(H) &=&j^{a}.  \label{ME.19}
\end{eqnarray}

Let $n^{a}=H^{a}/H$ \ be the unit vector in the direction of the magnetic
field. Geometrically $n^{a}$ is the unit tangent to the spacelike magnetic
field lines. The pair ($u^{a},n^{a})$ forms a double congruence. Maxwell
equations in terms of the irreducible parts defined by this double
congruence take a geometric form. The constraint equation (\ref{ME.16}) for
the magnetic field gives:%
\begin{equation*}
Hh_{b}^{a}n_{;a}^{b}+H_{,a}n^{a}=0.
\end{equation*}

But $h_{b}^{a}n_{;a}^{b}=p_{b}^{a}n_{;a}^{b}=\mathcal{E}$ \ where $\mathcal{E%
}$\ is the screen expansion of the magnetic field lines. Therefore 
\begin{equation}
\mathcal{E}=-(\ln H)^{\ast }  \label{ME.20}
\end{equation}%
where a \textquotedblleft *\textquotedblright\ over a symbol means covariant
differentiation wrt $n^{a}.$ From this equation we infer that the stronger
the magnetic field the denser the magnetic field lines on the screen space,
that is the greater is the magnetic flux through the screen space (as
expected). \ 

We examine now the propagation equation (\ref{ME.18}) of the magnetic field.
We have:

\begin{equation*}
\frac{\dot{H}}{H} n^{a} + h_{b}^{a} \overset{.}{n}^{b} =
u_{;b}^{a}n^{b}-\theta n^{a}.
\end{equation*}%
Contracting with $n^{a}$ and projecting with $p_{b}^{a}$ \ we get the pair
of equations:%
\begin{eqnarray}
(\ln H)^{\cdot } &=&\sigma _{ab}n^{a}n^{b}-\frac{2}{3}\theta  \label{ME.21}
\\
N^{a} &\equiv &p_{b}^{a}\mathcal{L}_{u}n^{b}=0.  \label{ME.22}
\end{eqnarray}

The first equation involves the change of the strength of the magnetic field
along the flow lines of the fluid i.e. the field $u^{a}.$ A kinematic
interpretation is that the vector $n^{a}$ is an eigenvector of the shear
with eigenvalue $(\ln H)^{\cdot }+\frac{2}{3}\theta .$

The second is the geometric condition that the magnetic field lines are
material lines in the fluid and correspond to the statement that the
magnetic field is \textquotedblleft frozen\textquotedblright\ along the
fluid. Physically this means that each particle of the fluid moves always on
the same magnetic field line.

Relation (\ref{ME.19}) \ due to (\ref{ME.24}) \ it is written as:%
\begin{equation}
\mathcal{R}^{a}=-\frac{H_{c}j^{c}}{2H^{3}}H^{a}  \label{ME.26}
\end{equation}%
therefore in the RMHD approximation the screen rotation of the magnetic
field lines vanishes iff the conduction current $j^{a}$ is normal to the
magnetic field.

Ohm's Law in its generalized form which includes the Hall current is written
(see \cite{dunn}):%
\begin{equation}
J^{a}=\rho u^{a}+\frac{1}{(1+\lambda ^{2}B^{c}B_{c})}\left[ kE^{a}+\lambda
k\eta ^{abcd}E_{b}u_{c}B_{d}+\lambda ^{2}k(E^{c}B_{c})B^{a}\right].
\label{Ohm.9}
\end{equation}

In the RMHD\ approximation we have that the spatial part $j^{a}$ \ of the
4-current $J^{a}$ vanishes, therefore $\mathcal{R}^{a}=0.$ Hence in a
perfectly conducting fluid for which generalized Ohm's law applies, the
magnetic field lines have zero rotation as measured by $u^{a}.$ Because (\ref%
{Ohm.9}) \ is not the most general form of Ohm's Law we shall assume in the
following $\mathcal{R}^{a}$ to be given by (\ref{ME.26}).

Summarizing we have that in the RMHD approximation Maxwell equations are:%
\begin{eqnarray}
\mathcal{E} &=&-(\ln H)^{\ast }  \label{ME.27} \\
\mathcal{R}^{a} &=&-\frac{H_{c}j^{c}}{2H^{3}}H^{a}  \label{ME.28} \\
e &=&2\omega ^{a}H_{a}  \label{ME.29} \\
(\ln H)^{\cdot } &=&\sigma _{ab}n^{a}n^{b}-\frac{2}{3}\theta  \label{ME.30}
\\
N^{a} &\equiv &p_{b}^{a}\mathcal{L}_{u}n^{b}=0.  \label{ME.31}
\end{eqnarray}%
Note that equation (\ref{ME.31}) can be written in the equivalent form:

\begin{equation}
p_{b}^{a}\overset{.}{n}^{b}=\left( p_{.c}^{a}\sigma _{b}^{c}+\omega
_{.b}^{a}\right) n^{b}.  \label{ME.32}
\end{equation}

\ These equations are general and independent of further simplifying
assumptions (e.g. symmetry assumptions) we might do, and hold in all cases.

\section{The field equations for the EMSF}

\label{section4}

The EMSF must satisfy three sets of equations: a) Maxwell equations, b)
Conservation laws and c) Einstein field equations. We have already given
Maxwell equations and the conservation equations.

Concerning Einstein field equations we shall consider their Lie derivative
along some characteristic direction of the EM fluid. The reason for this is
that we want to employ symmetry assumptions, that is equations of the form $%
\mathcal{L}_{\xi }M_{ab}=A_{ab}$ where $M_{ab}$ is a tensor computed in
terms of the metric (or the metric itself) and $A_{ab}$ is an arbitrary
tensor having the same symmetries as the $M_{ab}.$ Due to the form of
Einstein field equations we compute $\mathcal{L}_{\xi }R_{ab}$ \ in terms of 
$\mathcal{L}_{\xi }g_{ab}$ using various identities of Riemannian Geometry.
Then we impose the symmetry assumption by choosing a specific form for $%
A_{ab}.$ For example for a CKV $M_{ab}=g_{ab}$ and $A_{ab}=2\psi g_{ab}$
where $\psi (x^{a})$ \ is the conformal factor. \ Then we replace $\mathcal{L%
}_{\xi }R_{ab}$ in the Lie derivative of the field equations and we find the
field equations in a form that incorporates already the imposed geometric
symmetry assumption.

In a previous work on string fluids \cite{Tsamparlis M GRG 2006} we have
computed Einstein equations for a perfect string fluid and many types of
symmetries in the cases that the symmetry vector is either $\xi ^{a}=\xi
u^{a}$ or $\xi ^{a}=\xi n^{a}$. Therefore we could write straight away the
field equations in the case of an EMSF by simply specifying $n^{a}=H^{a}/H.$
Of course in this case the resulting equations will be supplemented by
Maxwell equations. In the following we recall briefly some important
intermediate steps in order to make the present work more readable and self
contained. Details can be found in \cite{Tsamparlis M GRG 2006}.

The Lie derivative of the Ricci tensor wrt a general time-like vector $\xi
^{a}=\xi u^{a}$ has been computed (see equation (3.9) in \cite{Saridakis
Tsamparlis 1991}) in terms of the standard dynamic variables $\mu
,p,q_{a},\pi _{ab}.$ By using the general expression, $L_{\xi }R_{ab}$ can
be written in terms of the perfect string fluid parameters $\rho ,q$. \ In a
similar way, the Lie derivative of the Ricci tensor along the spacelike
vector $\xi ^{a}=\xi n^{a}$ can be expressed in terms of the 1+1+2 dynamic
quantities.

Using Maxwell equations we show easily that the conservation equations (\ref%
{EqnSS.37}), (\ref{EqnSS.38}) and \eqref{EqnSS.39} (which must also be
satisfied in all cases) are simplified as follows:

\begin{eqnarray}
\overset{.}{\rho }-(\rho +p_{\perp })(\ln H)^{\cdot } &=&0  \label{MC.9} \\
\overset{\ast }{\rho }-(\rho +p_{\perp })(\ln H)^{\ast } &=&0  \label{MC.10}
\\
p_{a}^{b}\left[ p_{\perp }{}_{,b}+\lambda HH_{,b}+(\rho +p_{\perp }+\lambda
H^{2})(\overset{.}{u}_{b}-\overset{\ast }{n}_{b})\right] &=&0.  \label{MC.11}
\end{eqnarray}%
Concerning the Einstein filed equations we have from\cite{Saridakis
Tsamparlis 1991} and \cite{Tsamparlis M GRG 2006} the results:

\begin{eqnarray}
\frac{1}{\xi }L_{\xi }R_{ab} &=&\left[ \overset{.}{q}+2(q-\Lambda )(\ln \xi
)^{\cdot }\right] u_{a}u_{b}+2(q-\Lambda )\left[ \overset{.}{u}_{c}-(\ln \xi
)_{,c}\right] u_{(a}h_{b)}^{c} +  \notag \\
&&+\frac{1}{3}\left[ 2\overset{.}{\mu }-\overset{.}{q}+\frac{2}{3}(2\mu
-q+3\Lambda )\theta \right] h_{ab} +  \notag \\
&&+\left[ \overset{.}{\mu }+\overset{.}{q}+\frac{2}{3}(\mu +q)\theta \right]
\left( \frac{1}{3}h_{ab}-n_{a}n_{b} \right) + \frac{2}{3}(2\mu -q+3\Lambda
)\sigma_{ab} +  \notag \\
&& + 2 (\mu +q) \left(\frac{1}{3} h_{cd} - n_{c} n_{d} \right)
\delta_{(a}^{d}(\omega_{.b)}^{c}+\sigma _{.b)}^{c}) -  \notag \\
&&-2(\mu +q)\overset{.}{n}_{d}h_{(a}^{d}n_{b)}.  \label{N.33a}
\end{eqnarray}

\begin{eqnarray}
\frac{1}{\xi }L_{\xi }R_{ab} &=&\left[ \overset{\ast }{q}+2(q-\Lambda )%
\overset{.}{u}_{c}n^{c}\right] u_{a}u_{b}-2(q-\Lambda )\left[ \overset{\ast }%
{u}_{c}n_{c}-(\ln \xi )^{\cdot }\right] u_{(a}n_{b)} -  \notag \\
&&-\left[ \overset{\ast }{q}+2(q-\Lambda )(\ln \xi )^{\ast }\right]
n_{a}n_{b} -  \notag \\
&&-2\left[ (\mu +\Lambda )N_{c}+2(q-\Lambda )\omega _{dc}n^{d}\right]
u_{(a}p_{b)}^{c} -  \notag \\
&&-2(q-\Lambda )p_{c}^{d}\left[ \overset{\ast }{n}_{d}+(\ln \xi )_{,d}\right]
n_{(a}p_{b)}^{c}+\left[ \overset{\ast }{\mu }+(\mu +\Lambda )\mathcal{E}%
\right] p_{ab} +  \notag \\
&&+2(\mu +\Lambda )\mathcal{S}_{ab}.  \label{N.34a}
\end{eqnarray}

Expressions (\ref{N.33a}) and (\ref{N.34a}) are general and hold for \textit{%
all} collineations and all perfect string fluids. For each type of
collineation the lhs of the expressions (\ref{N.33a}) and (\ref{N.34a})
simplifies accordingly and for each specific perfect string fluid the rhs is
simplified the same. Equating the two parts one finds immediately Einstein
field equations for the specific string fluid considered and the specific
symmetry assumed.

In the case of the EMSF we have $\mu = \rho + \frac{1}{2}\lambda H^2$ and $q
= p_{\perp} + \frac{1}{2} \lambda H^2$. The result applies to all
collineations concerning the EMSF.

For easy reference we collect below the results of the calculations :

Maxwell equations:

\begin{eqnarray}
N^{a} &=&0 \iff p_{b}^{a}\overset{.}{n}^{b} = \left( p_{.c}^{a}\sigma
_{b}^{c}+\omega _{.b}^{a}\right) n^{b} \\
\mathcal{E} &\mathcal{=}&-(\ln H)^{\ast } \\
\sigma _{ab}n^{a}n^{b}-\frac{2}{3}\theta &=& \left( \ln H\right)^{\cdot} \\
e &=& 2 \omega^a H_a, \quad \mathcal{R}^a = - \frac{H_c j^c}{2H^3} H^a
\end{eqnarray}

Conservation equations:

\begin{eqnarray}
\overset{.}{\rho }-(\rho +p_{\perp })(\ln H)^{\cdot } &=&0 \\
\overset{\ast }{\rho }-(\rho +p_{\perp })(\ln H)^{\ast } &=&0 \\
p_{a}^{b}\left[ p_{\perp }{}_{,b}+\lambda HH_{,b}+(\rho +p_{\perp }+\lambda
H^{2})(\overset{.}{u}_{b}-\overset{\ast }{n}_{b})\right] &=&0.
\end{eqnarray}

Einstein filed equations:

\begin{eqnarray}
\frac{1}{\xi }L_{\xi }R_{ab} &=&\left[ \overset{.}{p_{\perp }}+\lambda H%
\overset{.}{H}+2\left( p_{\perp }+\frac{1}{2}\lambda H^{2}-\Lambda \right)
(\ln \xi )^{\cdot }\right] u_{a}u_{b}+2\left( p_{\perp }+\frac{1}{2}\lambda
H^{2}-\Lambda \right) \left[ \overset{.}{u}_{c}-(\ln \xi )_{,c}\right]
u_{(a}h_{b)}^{c}+  \notag \\
&&+\frac{1}{3}\left[ 2\overset{.}{\rho }-\overset{.}{p}_{\perp }+\lambda H%
\overset{.}{H}+\frac{2}{3}\left( 2\rho -p_{\perp }+\frac{1}{2}\lambda
H^{2}+3\Lambda \right) \theta \right] h_{ab}+  \notag \\
&&+\left[ \overset{.}{\rho }+\overset{.}{p}_{\perp }+2\lambda H\overset{.}{H}%
+\frac{2}{3}(\rho +p_{\perp }+\lambda H^{2})\theta \right] \left( \frac{1}{3}%
h_{ab}-n_{a}n_{b}\right) +\frac{2}{3}\left( 2\rho -p_{\perp }+\frac{1}{2}%
\lambda H^{2}+3\Lambda \right) \sigma _{ab}+  \notag \\
&&+2(\rho +p_{\perp }+\lambda H^{2})\left( \frac{1}{3}h_{cd}-n_{c}n_{d}%
\right) \delta _{(a}^{d}(\omega _{.b)}^{c}+\sigma _{.b)}^{c})-  \notag \\
&&-2(\rho +p_{\perp }+\lambda H^{2})\overset{.}{n}_{d}h_{(a}^{d}n_{b)}.
\label{N.35a}
\end{eqnarray}

\begin{eqnarray}
\frac{1}{\xi }L_{\xi }R_{ab} &=&\left[ \overset{\ast }{p_{\perp }}+\lambda H%
\overset{\ast }{H}+2\left( p_{\perp }+\frac{1}{2}\lambda H^{2}-\Lambda
\right) \overset{.}{u}_{c}n^{c}\right] u_{a}u_{b}-2(p_{\perp }+\frac{1}{2}%
\lambda H^{2}-\Lambda )\left[ \overset{\ast }{u}_{c}n^{c}-(\ln \xi )^{\cdot }%
\right] u_{(a}n_{b)}-  \notag \\
&&-\left[ \overset{\ast }{p_{\perp }}+\lambda H\overset{\ast }{H}+2\left(
p_{\perp }+\frac{1}{2}\lambda H^{2}-\Lambda \right) (\ln \xi )^{\ast }\right]
n_{a}n_{b}-  \notag \\
&&-4\left( p_{\perp }+\frac{1}{2}\lambda H^{2}-\Lambda \right) \omega
_{dc}n^{d}u_{(a}p_{b)}^{c}-  \notag \\
&&-2\left( p_{\perp }+\frac{1}{2}\lambda H^{2}-\Lambda \right) p_{c}^{d}%
\left[ \overset{\ast }{n}_{d}+(\ln \xi )_{,d}\right] n_{(a}p_{b)}^{c}+\left[ 
\overset{\ast }{\rho }+\lambda H\overset{\ast }{H}+\left( \rho +\frac{1}{2}%
\lambda H^{2}+\Lambda \right) \mathcal{E}\right] p_{ab}+  \notag \\
&&+2\left( \rho +\frac{1}{2}\lambda H^{2}+\Lambda \right) \mathcal{S}_{ab}.
\label{N.36a}
\end{eqnarray}

\textbf{At this point recall that for an arbitrary vector field }$\xi ^{a}$%
\textbf{\ we can always select }$n^{a}$\textbf{\ such that }%
\begin{equation*}
\xi ^{a}=\xi \left( u\right) u^{a}+\xi \left( n\right) n^{a}
\end{equation*}%
\textbf{to be always true. Hence, the Lie derivative }$L_{\xi }M_{ab}$%
\textbf{\ can be always expressed as }%
\begin{equation*}
\mathcal{L}_{\xi }M_{ab}=\mathcal{L}_{\xi \left( u\right) u^{a}}M_{ab}+%
\mathcal{L}_{\xi \left( n\right) n^{a}}M_{ab}
\end{equation*}%
\textbf{which means that we have two components. We continue by studying the
special cases where }$\xi \left( n\right) =0$\textbf{\ and }$\xi ^{a}$%
\textbf{\ is timelike, and }$\xi \left( u\right) =0$\textbf{\ and }$\xi ^{a}$%
\textbf{\ is spacelike.}

\section{The EMSF admitting a timelike CKV $\protect\xi^{a} = \protect\xi %
u^{a} \enskip (\protect\xi > 0)$}

\label{section5}

As it has mentioned there are two types of equations constraining the
evolutions of a gravitational system which admits a symmetry, that is the
kinematic conditions and dynamic equations.

The kinematic conditions are equations among the kinematic variables of the
gravitational system which result from geometric identities and additional
geometric assumptions (such as symmetries). The dynamic equations do not
necessarily inherit the kinematic symmetries of the system. In the following
we consider two types of symmetries (a) CKVs defined by timelike vectors $%
\xi ^{a}=\xi u^{a}$ $(\xi > 0)$; and (b) CKVs defined by the spacelike
vectors $\xi ^{a}=\xi n^{a}$ $(\xi > 0).$

\subsection{The case of a timelike CKV $\protect\xi^{a} = \protect\xi u^{a} 
\enskip (\protect\xi > 0)$}

We look first on the kinematic implications of the assumed symmetry and then
on the dynamical ones.

\subsection{The kinematic implications}

From previous works we have the following kinematic conditions for a CKV $%
\xi ^{a}=\xi u^{a}$ \cite{Noris Green P Davis 1977}.

\begin{proposition}
\label{Timelike CKV Kinematics.1}A fluid space-time $u^{a}$ admits a CKV $%
\xi ^{a}=\xi u^{a}$ iff:

\begin{enumerate}
\item $\sigma _{ab}=0$

\item $\overset{.}{u}_{a}=(\ln \xi )_{,a}+\frac{1}{3}\theta u_{a}$ where $%
\sigma _{ab},\theta $ and $\overset{.}{u}^{a}$ are, respectively, the shear,
expansion and acceleration of the timelike congruence generated by $u^{a}.$
The conformal factor $\psi =\frac{1}{3}\xi \theta = \dot{\xi}$.
\end{enumerate}
\end{proposition}

The conditions imposed by Proposition \ref{Timelike CKV Kinematics.1}
supplement Maxwell equations and simplify the conservation equations.
Because $\sigma _{ab}=0$ the \textquotedblleft energy\textquotedblright\
conservation equation (\ref{MC.9}) gives: 
\begin{equation}
\overset{.}{\rho }+\frac{2}{3}(\rho +p_{\perp })\theta =0.  \label{EqnSS.40}
\end{equation}

Equation (\ref{MC.10}) remains the same and \ equation (\ref{MC.11}) becomes:%
\begin{equation}
p_{a}^{b}\left[ p_{\perp }{}_{,b}+\lambda HH_{,b}+(\rho +p_{\perp }+\lambda
H^{2})((\ln \xi )_{,b}-\overset{\ast }{n}_{b})\right] =0.  \label{EqnSS.41}
\end{equation}

Eventually the conservation equations are equations (\ref{MC.10}), (\ref%
{EqnSS.40}), (\ref{EqnSS.41}).

\subsection{The dynamic implications}

For a CKV we have the identity:%
\begin{equation*}
L_{\xi }R_{ab}=-2\psi _{;ab}-g_{ab}\square \psi.
\end{equation*}
The 1+3 decomposition of $\psi_{;ab}$ wrt $u^{a}$ is (note that $\psi_{;ab}
= \psi_{;ba}$):

\begin{equation}
\psi _{;ab}=\lambda _{\psi }u_{a}u_{b}+p_{\psi }h_{ab}+2q_{\psi
(a}u_{b)}+\pi _{\psi ab}  \label{N.5}
\end{equation}%
where:%
\begin{equation}
\lambda _{\psi }=\psi _{;ab}u^{a}u^{b},\;p_{\psi }=\frac{1}{3}\psi
_{;ab}h^{ab},\;q_{\psi a}=-\psi _{;bc}h_{a}^{b}u^{c},\;\pi _{\psi ab}=
\left( h_{a}^{r}h_{b}^{s}-\frac{1}{3}h_{ab}h^{rs} \right) \psi_{;rs}.
\label{N.6}
\end{equation}
We also compute: 
\begin{equation}
\square \psi =\psi _{;ab}g^{ab}=-\lambda _{\psi }+3p_{\psi }.  \label{N6.a}
\end{equation}

Therefore for a CKV $\xi ^{a}=\xi u^{a}$ we have that:%
\begin{equation}
L_{\xi }R_{ab}=-\left[ 3(\lambda _{\psi }-p_{\psi })u_{a}u_{b}-(\lambda
_{\psi }-5p_{\psi })h_{ab}+4q_{\psi (a}u_{b)}+2\pi _{\psi ab}\right].
\label{N6.b}
\end{equation}%
Using the kinematic conditions and the conservation equations, the rhs of
equation \eqref{N.35a} simplifies as follows\footnote{%
It is easy to show (use Maxwell equations in RMHD approximation) that $%
\overset{.}{n}_{d}h_{(a}^{d}n_{b)}+n_{c}n_{d}\delta _{(a}^{d}\omega
_{.b)}^{c}=0$}:

\begin{eqnarray}
\frac{1}{\xi }L_{\xi }R_{ab} &=&\left[ \overset{.}{p_{\perp }}+\frac{1}{2}%
\lambda H\overset{.}{H}+2(p_{\perp }-\Lambda )\frac{1}{3}\theta \right]
u_{a}u_{b}-\frac{1}{3}\left[ \overset{.}{p}_{\perp }+2(p_{\perp }-\Lambda
)\theta -\frac{1}{2}\lambda H\overset{.}{H}\right] h_{ab}  \notag \\
&&+ \left[ \overset{.}{p}_{\perp }+\lambda H\overset{.}{H}\right] \left( 
\frac{1}{3} h_{ab}-n_{a}n_{b} \right).  \label{N6.c}
\end{eqnarray}

From (\ref{N6.b}) and (\ref{N6.c}) we find that the field equations for an
EM\ string fluid admitting the CKV $\xi ^{a}=\xi u^{a}$ are 
\begin{eqnarray*}
&& \left[ \overset{.}{p_{\perp }}+\frac{1}{2}\lambda H\overset{.}{H}%
+2(p_{\perp }-\Lambda )\frac{1}{3}\theta \right] u_{a}u_{b}-\frac{1}{3}\left[
\overset{.}{p}_{\perp }+2(p_{\perp }-\Lambda )\theta -\frac{1}{2}\lambda H%
\overset{.}{H}\right] h_{ab} \\
&&+\left[ \overset{.}{p}_{\perp }+\lambda H\overset{.}{H}\right] \left( 
\frac{1}{3} h_{ab}-n_{a}n_{b} \right) \\
&=&-\frac{1}{\xi }\left[ 3(\lambda _{\psi }-p_{\psi
})u_{a}u_{b}-(\lambda_{\psi } - 5 p_{\psi }) h_{ab} + 4 q_{\psi(a} u_{b)} +
2 \pi_{\psi ab}\right]
\end{eqnarray*}

This relation implies the field equations\footnote{%
The same equations are found from \cite{Tsamparlis M GRG 2006} where the
field equations were:
\par
{}%
\begin{eqnarray*}
\overset{.}{q} &=& - \frac{3}{\xi} (p_{\psi} + \lambda_{\psi }) \\
(q - \Lambda) \theta &=& \frac{9}{\xi} p_{\psi} \\
3(p_{\psi }+\lambda _{\psi })\left( \frac{1}{3}h_{ab}-n_{a}n_{b} \right)
&=&2\pi _{\psi ab}. \\
q_{\psi }^{a} &=&0
\end{eqnarray*}%
}:

\begin{eqnarray}
\overset{.}{p_{\perp }}+\lambda H\overset{.}{H} &=& - \frac{3}{\xi} (p_{\psi
}+\lambda_{\psi })  \label{EqnSS.45} \\
\left( p_{\perp }+\frac{1}{2}\lambda H^{2}-\Lambda \right) \theta &=& \frac{9%
}{\xi} p_{\psi }  \label{EqnSS.46} \\
3(p_{\psi }+\lambda _{\psi }) \left( \frac{1}{3}h_{ab}-n_{a}n_{b} \right)
&=& 2 \pi_{\psi ab}  \label{EqnSS.47} \\
q_{\psi }^{a} &=&0.  \label{EqnSS.48}
\end{eqnarray}

\bigskip Eqn (\ref{EqnSS.45}) using also $(ln\xi )^{\cdot }=\frac{1}{3}%
\theta $ can be written :%
\begin{eqnarray}
\left( p_{\perp }+\frac{1}{2}\lambda H^{2}-\Lambda \right)^{\cdot } &=& -
\left( p_{\perp }+\frac{1}{2}\lambda H^{2}-\Lambda \right)(ln\xi )^{\cdot
}-3\lambda _{\psi } \implies  \notag \\
\lambda_{\psi } &=&-\frac{1}{3}\left[ \left( p_{\perp }+\frac{1}{2}\lambda
H^{2}-\Lambda \right) \xi \right] ^{\cdot }  \label{EqnSS.49}
\end{eqnarray}

and (\ref{EqnSS.46}):%
\begin{equation}
p_{\psi }=\frac{1}{3} \left( p_{\perp }+\frac{1}{2}\lambda H^{2}-\Lambda
\right) \dot{\xi}.  \label{EqnSS.50}
\end{equation}

The final set of equations which results from the assumption that the EMSF\
admits the CKV $\xi ^{a}=\xi u^{a}$ is the following:

Geometric implications:%
\begin{eqnarray}
\sigma _{ab} &=&0 \\
\overset{.}{u}_{a} &=&(\ln \xi )_{,a}+\frac{1}{3}\theta u_{a}
\end{eqnarray}

Maxwell equations:

\begin{eqnarray}
N^{a} &=&0  \label{EqnSS.51} \\
\mathcal{E} &\mathcal{=}&-(\ln H)^{\ast }  \label{EqnSS.52} \\
p_{b}^{a}\overset{.}{n}^{b} &=&\omega _{.b}^{a}n^{b}  \label{EqnSS.53} \\
-\frac{2}{3}\theta &=&(\ln H)^{\cdot }  \label{EqnSS.54}
\end{eqnarray}

Conservation equations:

\begin{eqnarray}
\overset{.}{\rho }-(\rho +p_{\perp })(\ln H)^{\cdot } &=&0  \label{EqnSS.55}
\\
\overset{\ast }{\rho }-(\rho +p_{\perp })(\ln H)^{\ast } &=&0
\label{EqnSS.56} \\
p_{a}^{b}\left[ p_{\perp }{}_{,b}+\lambda HH_{,b}+(\rho +p_{\perp }+\lambda
H^{2})((\ln \xi )_{,b}-\overset{\ast }{n}_{b})\right] &=&0.  \label{EqnSS.57}
\end{eqnarray}

Gravitational Field equations:

\begin{eqnarray}
\lambda _{\psi } &=&-\frac{1}{3}\left[ \left( p_{\perp }+\frac{1}{2}\lambda
H^{2}-\Lambda \right) \xi \right] ^{\cdot }  \label{EqnSS.58} \\
p_{\psi } &=&\frac{1}{3} \left( p_{\perp }+\frac{1}{2}\lambda H^{2}-\Lambda
\right) \dot{\xi}  \label{EqnSS.59} \\
2\pi_{\psi ab} &=& - \xi \left( p_{\perp }+\frac{1}{2}\lambda H^{2}-\Lambda
\right)^{\cdot } \left( \frac{1}{3}h_{ab}-n_{a}n_{b} \right)
\label{EqnSS.60} \\
q_{\psi }^{a} &=&0.  \label{EqnSS.61}
\end{eqnarray}

This is the complete set of equations which must be satisfied by the various
variables (geometric, kinematic and dynamic) of an EMSF which admits the CKV 
$\xi ^{a}=\xi u^{a}.$

\subsection{The case of an EMSF admitting a timelike CKV $\protect\xi ^{a}=%
\protect\xi u^{a}$ in the FRW spacetime}

We apply the results of the last section to \ the case of the FRW spacetime.
The FRW spacetime has metric (in conformal coordinates):%
\begin{equation}
ds^{2}=R^{2}(\tau )\left[ -d\tau ^{2}+U^{2}(x^{\mu })d\sigma _{E}^{2}\right]
\label{FRW.1}
\end{equation}%
where the function $U^{2}(x^{\mu })=\left( 1+\frac{k}{4}\mathbf{x}\cdot 
\mathbf{x}\right) ^{-1}$ and $k=0,\pm 1.$ This metric admits the gradient
CKV $\partial _{\tau }$ whose conformal factor is $\psi = \frac{dR}{d\tau}$.
We define the timelike unit vector $u^{a}=\frac{1}{R}\partial _{\tau }.$ If
we define the new coordinate $t$ by the requirement:%
\begin{equation}
d\tau =\frac{1}{R(t)}dt  \label{FRW.2}
\end{equation}%
then the metric is written:%
\begin{equation}
ds^{2}=-dt^{2}+R^{2}(t)U^{2}(x^{\mu })d\sigma _{E}^{2}  \label{FRW.3}
\end{equation}%
and the unit vector $u^{a}=\partial _{t}.$ The conformal factor becomes: 
\begin{equation}
\psi =\frac{dR}{dt}\equiv \overset{.}{R}(t)  \label{FRW.4}
\end{equation}

Then\footnote{%
In the coordinates $\{t,x,y,z\}$ we have $g_{ab}$ $=$ $%
diag(-1,R^{2}U^{2},R^{2}U^{2},R^{2}U^{2})$,$~$that is, $g^{ab}$ $=$ $%
diag\left( -1,\frac{1}{R^{2}U^{2}},\frac{1}{R^{2}U^{2}},\frac{1}{R^{2}U^{2}}%
\right) ,$ where $u^{a}=(1,0,0,0)$, $u_{a}=(-1,0,0,0)$ and $\xi ^{a}=Ru^{a}$}
\begin{equation*}
\psi _{;ab}=\psi _{,ab}-\psi _{,c}\{_{ab}^{c}\}=\overset{...}{R}\delta
_{a}^{0}\delta _{b}^{0}-\psi _{,0}\{_{ab}^{0}\}=\overset{...}{R}u_{a}u_{b}-%
\frac{1}{2}\ddot{R}g_{ab,0}
\end{equation*}%
where 
\begin{equation*}
\{_{ab}^{0}\}=\frac{1}{2}g^{0c}(g_{ac,b}+g_{bc,a}-g_{ab,c})=\frac{1}{2}%
g^{00}(g_{a0,b}+g_{b0,a}-g_{ab,0})=\frac{1}{2}g_{ab,0}.
\end{equation*}%
For $\mu =1,2,3$ we find $\{_{\mu \mu }^{0}\}=\frac{1}{2}g_{\mu \mu
,0}=U^{2}R\dot{R}$ and $g_{\mu \mu ,0}=2U^{2}R\dot{R}$. From 1+3
decomposition over $\psi _{;ab}$ 
\begin{equation*}
\lambda _{\psi }=\overset{...}{R},\enskip p_{\psi }=-\frac{\dot{R}\ddot{R}}{R%
},\enskip q_{\psi }^{a}=0,\enskip\pi _{\psi ab}=-\frac{1}{2}\ddot{R}g_{ab,0}+%
\frac{\dot{R}\ddot{R}}{R}h_{ab}.
\end{equation*}%
Equation \eqref{EqnSS.59} gives ($\dot{R}\neq 0$) 
\begin{equation*}
\ddot{R}=-\frac{1}{3}\left( p_{\perp }+\frac{1}{2}\lambda H^{2}-\Lambda
\right) R.
\end{equation*}%
Then $p_{\perp }+\frac{1}{2}\lambda H^{2}-\Lambda =0$ is a possible
assumption which implies that $\ddot{R}=0$; thus $\psi =\dot{R}=const$ and $%
R(t)=bt+c$ for some real constants $b,c$. \vspace{20pt} %%%%%

We compute:%
\begin{equation*}
\psi _{;ab}=\overset{...}{R}\delta _{a}^{t}\delta _{b}^{t}
\end{equation*}
which implies:%
\begin{equation}
\lambda _{\psi }=\overset{...}{R}, \enskip p_{\psi }=0, \enskip q_{\psi a}, %
\enskip \pi _{\psi ab}=0.  \label{FRW.5}
\end{equation}

Field equation (\ref{EqnSS.50}) gives (as $\psi \neq const$ and $\dot{R}
\neq 0$):%
\begin{equation*}
p_{\perp }=-\frac{1}{2}\lambda H^{2}+\Lambda
\end{equation*}%
and (\ref{EqnSS.49}) $\lambda _{\psi }=\overset{...}{R}$ $=0.$ It follows
that:

1) $\partial _{\tau }=R(t)u^{a}$ is a special CKV or one of its
specializations (If $\psi =0\Rightarrow R(t)=const$ the spacetime reduces to
a Einstein space);

2) The conformal factor is $\psi =bt+c;\,$

3) The spacetime admits the special gradient CKV $\psi _{,a}=b\delta
_{a}^{t} $;

4) The scale factor $R(t)=\frac{1}{2}bt^{2}+ct+d.$

We work now with the rest of the conservation equations. Equation %
\eqref{EqnSS.55} using $p_{\perp }=-\frac{1}{2}\lambda H^{2}+\Lambda$ gives 
\begin{eqnarray}
\overset{.}{\mu }- \left( \mu -\frac{1}{2}\lambda H^{2}+\Lambda \right)(\ln
H)^{\cdot } &=& 0 \implies \overset{.}{\mu }-(\mu +\Lambda )(\ln H)^{\cdot }+%
\frac{1}{2}\lambda H^{2}(\ln H)^{\cdot } = 0 \implies  \notag \\
\frac{\overset{.}{\mu }}{\mu +\Lambda }-\frac{\overset{.}{H}}{H}+\frac{1}{2}%
\lambda \frac{H}{\mu +\Lambda }\overset{.}{H} &=& 0 \implies \left( \ln 
\frac{\mu +\Lambda }{H}\right) ^{\cdot }+\frac{1}{2}\lambda \frac{H}{\mu
+\Lambda }\overset{.}{H} = 0 \implies  \notag \\
\left( \frac{\mu +\Lambda }{H}\right) ^{\cdot }+\frac{1}{2}\lambda \overset{.%
}{H} &=& 0 \implies  \notag \\
\left( \mu +\frac{1}{2}\lambda H^{2}\right)^{\cdot } &=& 0  \label{FRW.6}
\end{eqnarray}

This equation means that the energy of the EM string\ fluid is constant
along the flow lines of the observers{\Large .}

Working similarly we show that equation (\ref{EqnSS.56}) becomes:%
\begin{equation}
\left( \mu +\frac{1}{2}\lambda H^{2}\right) ^{\ast }=0  \label{FRW.7}
\end{equation}%
and implies that the total energy of the magnetofluid is conserved along the
magnetic field lines. \ Both these results are compatible with:

a. The fact that the magnetic field lines are frozen along the flow lines of
the fluid (there is no relative motion of the two sets of lines) due to the
condition $N^{a}=0$

b. The dynamic equation $q_{\psi}^{a}=0$ \ i.e. there is no heat flux wrt
the observers $u^{a}.$

There remains equation (\ref{EqnSS.57}). Taking into account the fact that $%
\xi (t)$ (hence $p_{b}^{a}\xi _{,a}=0)$ we find that this equation becomes:%
\begin{equation*}
\left( \mu +\Lambda +\frac{1}{2}\lambda H^{2}\right) p_{a}^{b}\overset{\ast }%
{n^{a}} =0.
\end{equation*}

Because the total energy of the fluid (including the cosmological constant)
is considered to be positive this equation gives the condition:%
\begin{equation*}
p_{a}^{b}\overset{\ast }{n^{a}}=0.
\end{equation*}

This is a dynamical equation which involves the magnetic field only. However
we also have $\overset{\ast }{n^{a}}u_{a}=-\sigma _{ab}n^{a}n^{b}-\frac{%
\theta }{3}=0$. But since $\sigma _{ab}=0$ we find that $\overset{\ast }{%
n^{a}}u_{a}=-\frac{\theta }{3}$ which from $p_{a}^{b}\overset{\ast }{n^{a}}%
=0 $ gives 
\begin{equation*}
\overset{\ast }{n^{a}}=\frac{1}{3}\theta u^{a}=(lnR)^{\cdot }u^{a}.
\end{equation*}

Eventually we have that the magnetic field lines are carried along with the
fluid so that the total energy density (that is fluid energy and magnetic
field energy) remains constant. Furthermore the fluid does not heat.

The magnetic field lines are coplanar with the fluid lines but they are not
Lie transported along these lines except in the case of Minkowski spacetime.
Indeed from the condition $N^{a}=0$ we have $L_{u}n^{a}=au^{a}+bn^{a}$ where 
$a,b$ \ are quantities which have to be computed. From the definition of the
Lie derivative we have $L_{u}n^{a}=\overset{.}{n}^{a}-\overset{\ast }{u^{a}}$
therefore we have:%
\begin{equation*}
\overset{.}{n}^{a}-\overset{\ast }{u^{a}}=au^{a}+bn^{a}.
\end{equation*}
Contracting in turn with $u^{a},n^{a}$ we find $a=(\ln \xi )^{\ast },$ $b=-%
\frac{1}{3}\theta $ therefore:%
\begin{equation}
\mathcal{L}_{u}n^{a}=(\ln \xi )^{\ast }u^{a}-\frac{1}{3}\theta n^{a}
\end{equation}
which proves our assertion. From this relation it follows that $%
p_{b}^{a}L_{u}n^{a}=0$, that is $N^{a}=0$.

Concerning the magnetic field we have 
\begin{equation}
L_{u}H^{a}=\overset{.}{H}^{a}-\overset{\ast}{u^{a}} H = \overset{.}{H}%
n^{a}+H \mathcal{L}_{u}n^{a}=(\ln \xi )_{,b} H^b u^{a}-\theta H^{a}.
\end{equation}

\section{The EMSF in spacetimes admitting a spacelike CKV $\ \protect\xi %
^{a}=\protect\xi n^{a}$}

\label{section6}

We derive again the kinematic and the dynamic equations as we did for the
case of $\xi ^{a}=\xi u^{a}.$

\subsection{The Kinematic conditions of a spacelike CKV $\protect\xi ^{a}=%
\protect\xi n^{a}$}

For a double congruence $u^{a},n^{a}$ one has the kinematic quantities $%
\sigma _{ab,}\omega _{ab},$ $\theta ,$ $\overset{.}{u}_{a}$ for the timelike
congruence $u^{a}$ \ and the kinematic quantities $\mathcal{S}_{ab},\mathcal{%
R}_{ab},\mathcal{E}$ $\overset{.}{n}_{a}$, $\overset{\ast }{u}_{a}$ \ for
the spacelike congruence $n^{a}.$ Therefore the kinematic restrictions in
this case involve in general all nine quantities plus the parameters $\psi $
and $H_{ab}$ and their derivatives. To find the kinematic conditions
resulting from a collineation relative to a double congruence we need the
1+1+2 decomposition of $H_{ab}.$ To do that we consider the symmetry
defining equation and contract it to get:%
\begin{equation*}
\psi =\frac{\xi }{4}\left[ \mathcal{E}+(\ln \xi )^{\ast }-\overset{.}{n}%
^{c}u_{c}\right]
\end{equation*}

For the case of a CKV\ $\xi ^{a}=\xi n^{a}$ we find the following kinematic
conditions \cite{Tsamparlis Mason 1990}:

\begin{proposition}
\label{Spacelike CKV Kinematic}A fluid spacetime $u^{a}$ with a spacelike
congruence $n^{a}$ ($u^{a}n_{a}=0)$ admits the spacelike CKV\footnote{$\xi $
is not necessarily equal to $H!$} $\xi ^{a}=\xi n^{a}$ ($\xi >0)$ iff:%
\begin{eqnarray}
\mathcal{S}_{ab} &=&0  \label{KC.18} \\
\overset{.}{n}_{a}u^{a} &=&-\frac{1}{2}\mathcal{E}  \label{KC.19} \\
\overset{\ast }{n^{a}} &=&(\ln \xi )^{\cdot }u^{a}-p^{ab}(\ln \xi )_{,b}
\label{KC.20} \\
N_{a} &=&-2\omega _{ab}n^{b}.  \label{KC.21}
\end{eqnarray}
\end{proposition}

The conformal factor $\psi $ satisfies:%
\begin{equation}
\psi =\frac{1}{2}\xi \mathcal{E=}\overset{\ast }{\xi }.  \label{KC.22}
\end{equation}

Furthermore we can show the Lie derivatives \cite{Maartens Mason Tsamparlis
1985}:%
\begin{eqnarray}
L_{\xi }n^{a} &=&-\psi n^{a}  \label{KC.23} \\
L_{\xi }u^{a} &=&=-\psi u^{a}-\xi N^{a}.  \label{KC.24}
\end{eqnarray}
We note that:%
\begin{equation}
\mathcal{E=}(\ln \xi ^{2})^{\ast }.  \label{KC.24a}
\end{equation}

Also $\overset{\ast }{n^{a}}$ is the principal normal to the magnetic field
lines. \ We note that in general these lines are not straight lines. The
main results on the kinematics of a CKV $\xi ^{a}=\xi n^{a}$ are given in
the following Proposition (see Theorem 4.1. of \cite{Saridakis Tsamparlis
1991}):

\begin{proposition}
\label{Theorem 1}Let $\xi ^{a}=\xi n^{a}$ be a spatial conformal Killing
vector orthogonal to $u^{a}.$ Then $L_{\xi }n_{a}=\psi n_{a}.$ Furthermore
the following statements are equivalent:
\end{proposition}

\begin{enumerate}
\item $N^{a}=0$

\item $\omega ^{a}\parallel \xi ^{a}$ or $\omega ^{a}=0$

\item $L_{\xi }u_{a}=\psi u_{a}$

\item $L_{\xi }\omega _{ab}=\psi \omega _{ab}$

\item $L_{\xi }\sigma _{ab}=\psi \sigma _{ab}$

\item $L_{\xi }\overset{.}{u}_{a}=\psi ,_{a}+\overset{.}{\psi }u_{a}$

\item $L_{\xi }\theta =-\psi \theta +3\overset{.}{\psi }$
\end{enumerate}

We have the obvious identity:%
\begin{equation*}
\overset{.}{n}^{a}=-(\overset{.}{n}_{b}u^{b})u^{a}+p_{b}^{a}\overset{.}{n}%
^{b}.
\end{equation*}
We also have:%
\begin{eqnarray}
N^{a} &=&p_{b}^{a}(\mathcal{L}_{u}n^{b})=p_{b}^{a}(\overset{.}{n}^{b}-%
\overset{\ast }{u}^{b})=p_{b}^{a}\overset{.}{n}^{b}-p_{b}^{a}\overset{\ast }{%
u}^{b}  \notag \\
&=&p_{b}^{a}\overset{.}{n}^{b}-p_{b}^{a}\left( \sigma _{c}^{b}+\omega
_{c}^{b}\right) n^{c}=p_{b}^{a}\overset{.}{n}^{b}-p_{b}^{a}\sigma
_{c}^{b}n^{c}-\omega _{\;\;c}^{a}n^{c}  \notag \\
&=&p_{b}^{a}\overset{.}{n}^{b}-p_{b}^{a}\sigma _{c}^{b}n^{c}+\frac{1}{2}%
N^{a}\Rightarrow  \notag \\
p_{b}^{a}\overset{.}{n}^{b} &=&p_{b}^{a}\sigma _{c}^{b}n^{c}+\frac{1}{2}%
N^{a}.  \label{KC.24b}
\end{eqnarray}%
Using the symmetry equation we find:%
\begin{equation}
\overset{.}{n}^{a}=\frac{1}{2}\mathcal{E}u^{a}+p_{b}^{a}\sigma _{c}^{b}n^{c}+%
\frac{1}{2}N^{a}.  \label{KC.24c}
\end{equation}

\subsection{The dynamic conditions of a spacelike CKV $\protect\xi ^{a}=%
\protect\xi n^{a}$}

We have to consider three sets of equations i.e. Maxwell equations, the
conservation equations and the gravitational field equations.

\subsubsection{Maxwell equations}

The above results hold for any spacelike CKV and any string fluid. For the
particular case of the EMSF we have to supplement these equations with
Maxwell equations which are 
\begin{eqnarray}
N^{a} &=&0 \iff p_{b}^{a}\overset{.}{n}^{b}=p_{.c}^{a}\sigma _{b}^{c}n^{b}
\label{KC.25} \\
\mathcal{E} &\mathcal{=}&-(\ln H)^{\ast }  \label{KC.26} \\
\sigma _{ab}n^{a}n^{b}-\frac{2}{3}\theta &=&\left( \ln H\right) ^{\cdot }
\label{KC.27} \\
e &=&2\omega ^{a}H_{a},\text{ \ }\mathcal{R}^{a}=-\frac{j^{b}H_{b}}{2H^{3}}%
H^{a}.
\end{eqnarray}

Using $\mathcal{E}\mathcal{=}-(\ln H)^{\ast }$ and (\ref{KC.24a}) we find:%
\begin{equation*}
(\xi ^{2}H)^{\ast }=0
\end{equation*}

that is the quantity $\xi ^{2}H$ is constant along the magnetic field lines.

We also conclude that $\omega ^{a}\parallel H^{a}$ \ that is the magnetic
field congruence coincides with the vorticity congruence.

Using Maxwell equations we show the following important Proposition.

\begin{proposition}
$\xi ^{a}$ is a CKV of the screen metric $p_{ab}$ as well as of the 3-metric 
$h_{ab}$ with conformal factor $\psi =\frac{1}{2}\xi \mathcal{E}$ (the same
for both metrics).
\end{proposition}

Proof

In \cite{Tsamparlis M GRG 2006} it has been shown (see relations (26),(27))
that the following general relations/identities hold for the Lie derivatives
of the projection tensors $h_{ab}$ and \ $p_{ab}:$ 
\begin{eqnarray}
\frac{1}{\xi }L_{\xi }p_{ab} &=&2\left( \mathcal{S}_{ab}+\frac{1}{2}\mathcal{%
E}p_{ab}\right) -2u_{(a}N_{b)}  \label{KC.28a} \\
\frac{1}{\xi }L_{\xi }h_{ab} &=&2\left( \mathcal{S}_{ab}+\frac{1}{2}\mathcal{%
E}p_{ab}\right) -2u_{(a}N_{b)}+2(\ln \xi )_{,(a}n_{b)}+2\overset{\ast }{n}%
_{(a}n_{b)}.  \label{KC.28b}
\end{eqnarray}

From equation (\ref{KC.25}) and the kinematic condition (\ref{KC.18}) \
these equations reduce as follows: 
\begin{eqnarray}
\frac{1}{\xi }L_{\xi }p_{ab} &=&\mathcal{E}p_{ab}  \label{KC.28c} \\
\frac{1}{\xi }L_{\xi }h_{ab} &=&\mathcal{E}p_{ab}+2(\ln \xi )_{,(a}n_{b)}+2%
\overset{\ast }{n}_{(a}n_{b)}.  \label{KC.28d}
\end{eqnarray}%
It follows immediately that $\xi ^{a}$ is a CKV for the 2-metric \ $p_{ab}$
\ in the screen space with conformal factor $\frac{1}{2}\xi \mathcal{E}.$

To show that $\xi ^{a}$ is a CKV for the 3-metric $h_{ab}$ we 1+1+2
decompose $(\ln \xi )_{,a}$ in terms of the vectors $u^{a},n^{a}$ \ and find:%
\begin{equation*}
(\ln \xi )_{,a}=-(\ln \xi )^{\cdot }u_{a}+(\ln \xi )^{\ast }n_{a}+p^c_a(\ln
\xi )_{,c}.
\end{equation*}

From (\ref{KC.28d}) \ and (\ref{KC.20}) \ \ we have then:%
\begin{eqnarray*}
\frac{1}{\xi }L_{\xi }h_{ab} &=&\mathcal{E}p_{ab}+2(\ln \xi )_{,(a}n_{b)}+2%
\overset{\ast }{n}_{(a}n_{b)} \\
&=& \mathcal{E}p_{ab}+2\left[ \overset{\ast }{n}_{d}-(\ln \xi )^{\cdot
}u_{d}+(\ln \xi )^{\ast }n_{d}+ p^c_d (\ln \xi )_{,c}\right] \delta^d_{(a}
n_{b)} \\
&=& \mathcal{E}p_{ab}+2(\ln \xi )^{\ast }n_{a}n_{b}.
\end{eqnarray*}

But from (\ref{KC.22}) \ we have that $\mathcal{E=}2(\ln \xi )^{\ast }$
therefore:%
\begin{equation*}
\frac{1}{\xi }L_{\xi }h_{ab}=\mathcal{E(}p_{ab}+n_{a}n_{b})=\mathcal{E}h_{ab}
\end{equation*}

from which it follows that $\xi ^{a}$ is a CKV for the 3-metric $h_{ab}$
with conformal factor $\frac{1}{2}\xi \mathcal{E}.$ $\boxdot $

From (\ref{KC.24c}) \ we also have\footnote{%
One could possibly expect to get information on $\sigma _{bc}n^{b}n^{c}$from
this equation. But this is not so. Indeed by expanding $p_{b}^{a}$ we find:%
\begin{equation*}
\overset{.}{n}^{a}=-\frac{1}{2}(\ln H)^{\ast }u^{a}+\sigma
_{c}^{a}n^{c}-(\sigma _{bc}n^{b}n^{c})n^{a}
\end{equation*}%
\par
and we get no information on $\sigma _{bc}n^{b}n^{c}.$%
\par
{}}:%
\begin{equation*}
\overset{.}{n}^{a}=-\frac{1}{2}(\ln H)^{\ast }u^{a}+p_{b}^{a}\sigma
_{c}^{b}n^{c}.
\end{equation*}

\subsubsection{Conservation equations}

These equations are the same as before, that is we have: 
\begin{eqnarray}
\overset{.}{\mu }-(\mu +p_{\perp })(\ln H)^{\cdot } &=&0  \label{KC.29} \\
\overset{\ast }{\mu }-(\mu +p_{\perp })(\ln H)^{\ast } &=&0  \label{KC.30} \\
p_{a}^{b}\left[ p_{\perp }{}_{,b}+\lambda HH_{,b}+(\mu +p_{\perp }+\lambda
H^{2})(\overset{.}{u}_{b}-\overset{\ast }{n}_{b})\right] &=&0.  \label{KC.31}
\end{eqnarray}

\subsubsection{Gravitational field equations}

We use (\ref{N.36a}) \ to compute these equations. Of course we can also
take them directly from \cite{Tsamparlis M GRG 2006} but we prefer to derive
them here in order to make clear the methods we follow.

First we compute the $L_{\xi }R_{ab}.$We note that:%
\begin{equation}
p_{\psi }=\frac{1}{3}\psi _{;ab}h^{ab}=\frac{1}{3}\psi
_{;ab}(p^{ab}+n^{a}n^{b})=\frac{1}{3}(\gamma _{\psi }+a_{\psi }).
\label{KC.32}
\end{equation}%
We have:%
\begin{eqnarray}
L_{\xi }R_{ab} &=&-2\psi _{;ab}-g_{ab}\square \psi  \notag \\
&=&-2\left[ \lambda _{\psi }u_{a}u_{b}+2k_{\psi }u_{(a}n_{b)}+2\mathcal{S}%
_{\psi (a}u_{b)}+\gamma _{\psi }n_{a}n_{b}+2P_{\psi (a}n_{b)}+\frac{1}{2}%
\alpha _{\psi }p_{ab}+D_{\psi ab}\right] -  \notag \\
&&-(-\lambda _{\psi }+3p_{\psi })(-u_{a}u_{b}+n_{a}n_{b}+p_{ab})  \notag \\
&=&3(p_{\psi }-\lambda _{\psi })u_{a}u_{b}+(\lambda _{\psi }-3p_{\psi
}-2\gamma _{\psi })n_{a}n_{b}+(\lambda _{\psi }-3p_{\psi }-a_{\psi
})p_{ab}+rest.  \label{KC.33}
\end{eqnarray}

From (\ref{N.36a}) \ we get the following field equations (including
equations $k_{\psi} = 0$, $S_{\psi a} = 0$, $P_{\psi a} = 0$, $D_{\psi ab} =
0$ which result directly from the kinematic conditions over \eqref{N.36a}):%
\begin{eqnarray*}
\overset{\ast }{p_{\perp }}+\lambda H\overset{\ast }{H}+2 \left(p_{\perp }+%
\frac{1}{2}\lambda H^{2}-\Lambda \right)\frac{1}{2}\mathcal{E} &=&\frac{1}{%
\xi }3(p_{\psi }-\lambda _{\psi }) \\
\overset{\ast }{p_{\perp }}+\lambda H\overset{\ast }{H+}2 \left(p_{\perp }+%
\frac{1}{2}\lambda H^{2}-\Lambda \right)\frac{1}{2}\mathcal{E} &=&-\frac{1}{%
\xi }(\lambda _{\psi }-3p_{\psi }-2\gamma _{\psi }) \\
\overset{\ast }{\mu }+\lambda H\overset{\ast }{H}+ \left( \mu +\frac{1}{2}%
\lambda H^{2}+\Lambda \right) \mathcal{E} &\mathcal{=}&\frac{1}{\xi }%
(\lambda _{\psi }-3p_{\psi }-a_{\psi }).
\end{eqnarray*}

Using equation (\ref{KC.26}) to replace $\mathcal{E}$ in terms of $(\ln
H)^{\ast }$ we note that the first two equations have identical lhs and they
result in the two \ equations:%
\begin{eqnarray*}
\overset{\ast }{p_{\perp }}- \left( p_{\perp }-\frac{1}{2}\lambda
H^{2}-\Lambda \right)(\ln H)^{\ast } &=&\frac{1}{\xi }3(p_{\psi }-\lambda
_{\psi }) \\
\lambda _{\psi } &=&-\gamma _{\psi }.
\end{eqnarray*}

The last equation is written:%
\begin{equation*}
\overset{\ast }{\mu }- \left(\mu -\frac{1}{2}\lambda H^{2}+\Lambda
\right)(\ln H)^{\ast }\mathcal{=}\frac{1}{\xi }(\lambda _{\psi }-3p_{\psi
}-a_{\psi }).
\end{equation*}%
Using the conservation equation (\ref{KC.30}) and equation \eqref{KC.32} \
we find: \ \ \ \ \ 
\begin{equation*}
\left( p_{\perp }+\frac{1}{2}\lambda H^{2}-\Lambda \right)(\ln H)^{\ast }=-%
\frac{2}{\xi }(\gamma _{\psi }+a_{\psi }).
\end{equation*}
Finally we have that the field equations in the case of a spacelike vector $%
\xi ^{a}=\xi n^{a}$ are:%
\begin{eqnarray}
\overset{\ast }{p_{\perp }}+\lambda H\overset{\ast }{H} &=&\frac{1}{\xi }%
(2\gamma _{\psi }-a_{\psi })  \label{KC.34} \\
\left( p_{\perp }+\frac{1}{2}\lambda H^{2}-\Lambda \right)(\ln H)^{\ast }
&=&-\frac{2}{\xi }(\gamma _{\psi }+a_{\psi }).  \label{KC.35}
\end{eqnarray}
where:%
\begin{equation}
\psi_{;ab}=-\gamma _{\psi }(u_{a}u_{b}-n_{a}n_{b})+\frac{1}{2}\alpha _{\psi
}p_{ab}.  \label{KC.36}
\end{equation}

We see that $\psi _{;ab}$ is the energy momentum tensor or equivalently the
Ricci tensor of a string fluid with $\rho =-\gamma _{\psi }$ \ and $q=\frac{1%
}{2}\alpha _{\psi }$ or vice versa. \ Obviously one can make many scenarios
with this observation.

The result we found coincides with the one we found in \cite{Tsamparlis M
GRG 2006} on strings.

From equations (\ref{KC.34}) \ and (\ref{KC.35}) \ ones shows easily that:%
\begin{equation}
\left[ \left(p_{\perp }+\frac{1}{2}\lambda H^{2}-\Lambda \right) H\right]
^{\ast }=-\frac{3H}{\xi }\alpha _{\psi } .  \label{KC.37}
\end{equation}

This equation shows that if $\alpha _{\psi }=p^{ab}\psi _{;ab}=0$ \ then the
quantity $(p_{\perp }+\frac{1}{2}\lambda H^{2}-\Lambda )H$ \ is constant
along the magnetic field lines.

The constraint equation\footnote{%
This equation follows form the identity $(R^{ab}\xi _{b})_{;a}=-3\square
\psi $ \ which holds for all CKVs.} for a general anisotropic fluid of the
form we consider is:%
\begin{equation}
(\mu -2p_{\perp }+p_{\shortparallel }+2\Lambda )\psi =2( 2 \lambda_{\psi
}-a_{\psi }).  \label{KC.38}
\end{equation}%
Setting $\mu =-p_{\shortparallel }$ and $p_{\perp }=p_{\perp }+\frac{1}{2}%
\lambda H^{2}$ \ we obtain the EMSF. In this case equation (\ref{KC.38})
becomes:%
\begin{equation}
(p_{\perp }+\frac{1}{2}\lambda H^{2}-\Lambda )\psi = a_{\psi } + 2
\gamma_{\psi }.  \label{KC.39}
\end{equation}
But $\psi =\frac{1}{2}\xi \mathcal{E}=-\frac{1}{2}\xi (\ln H)^{\ast }$
therefore we obtain the same result$.$

We collect the above results in the following Proposition.

\begin{proposition}
\label{Spacelike CKV}An EMSF spacetime admits a CKV of the form $\xi
^{a}=\xi n^{a}$ where $n^{a}=H^{a}/H$ \ iff the following system of
equations is satisfied:%
\begin{eqnarray}
\overset{.}{\mu }-(\mu +p_{\perp })(\ln H)^{\cdot } &=&0  \label{KC.40} \\
\overset{\ast }{\mu }-(\mu +p_{\perp })(\ln H)^{\ast } &=&0  \label{KC.41} \\
p_{a}^{b}\left[ \left( p_{\perp }+\frac{1}{2}\lambda H^{2}-\Lambda
\right)_{,b}+(\mu +p_{\perp }+\lambda H^{2})(\overset{.}{u}_{b}-\overset{%
\ast }{n}_{b})\right] &=&0  \label{KC.42} \\
\psi _{;ab} &=&-\gamma _{\psi }(u_{a}u_{b}-n_{a}n_{b})+\frac{1}{2}\alpha
_{\psi }p_{ab}  \label{KC.43} \\
\left[ \left( p_{\perp }+\frac{1}{2}\lambda H^{2}-\Lambda \right) H \right]%
^{\ast } &=&-\frac{3H}{\xi }\alpha _{\psi }  \label{KC.44} \\
\left( p_{\perp }+\frac{1}{2}\lambda H^{2}-\Lambda \right)(\ln H)^{\ast }
&=&-\frac{2}{\xi }(\gamma _{\psi }+a_{\psi })  \label{KC.45} \\
\mathcal{S}_{ab} &=&0,\text{ \ }\mathcal{E}=-(\ln H)^{\ast }  \label{KC.46}
\\
\overset{.}{n}^{a} &=&-\frac{1}{2}(\ln H)^{\ast }u^{a}+p_{b}^{a}\sigma
_{c}^{b}n^{c}  \label{KC.47} \\
\overset{\ast }{n}^{a} &=&(\ln \xi )^{\ast }n^{a}-(\ln \xi )^{,a}
\label{KC.48} \\
N_{a} &=&0  \label{KC.49} \\
\sigma _{ab}n^{a}n^{b}-\frac{2}{3}\theta &=&\left( \ln H\right) ^{\cdot }
\label{KC.49a} \\
e &=&2\omega^a H_a,\text{ \ }\mathcal{R}^{a}=-\frac{j^{b}H_{b}}{2H^{3}}H^{a}.
\label{KC49.b}
\end{eqnarray}%
Furthermore the rotation $\omega ^{a}$ is either parallel to $H^{a}$ \ or
vanishes \ and $\psi =\frac{1}{2}\xi \mathcal{E}=-\frac{1}{2}\xi (\ln
H)^{\ast }.$
\end{proposition}

One important result is that if the vorticity vanishes then the same must be
true for the charge density and conversely. This is a restriction of
physical nature resulting from a geometrical symmetry assumption.

A\ CKV\ for which $\psi _{;ab}=0$ is called a special CKV. Coley and Tupper 
\cite{Coley Tupper (1989)} have shown that if an anisotropic fluid
space-time admits a proper special CKV $\xi ^{a}=\xi n^{a}$ then (assuming $%
\Lambda =0)$:%
\begin{equation}
\mu =-p_{\parallel }=\frac{1}{2}R,\;p_{\perp }=0  \label{KC.50}
\end{equation}%
where $R$ is the Ricci scalar. From Einstein field equations it follows that
for this case $T_{ab}$ is of the form:%
\begin{equation}
T_{ab}=\frac{1}{2}R(u_{a}u_{b}-n_{a}n_{b}).  \label{KC.51}
\end{equation}

For the case of a string fluid this result gives $\rho =\frac{1}{2}R$ and $%
q=0.$ Obviously $R\neq 0$ otherwise we do not have a fluid at all. Let us
check if our results are compatible with this general result.

From equation (\ref{KC.45}) \ \ assuming $\overset{\ast }{H}\neq 0$ we have:%
\begin{equation}
p_{\perp }+\frac{1}{2}\lambda H^{2}-\Lambda =0  \label{KC.52}
\end{equation}
which gives immediately from (\ref{EqnSS.25}):%
\begin{equation*}
R_{ab}= \left( \mu +\frac{1}{2}\lambda H^{2}+\Lambda \right) p_{ab}
\end{equation*}
from which follows:%
\begin{equation}
R=2 \left( \mu +\frac{1}{2}\lambda H^{2}+\Lambda \right).  \label{KC.52a}
\end{equation}
Therefore:%
\begin{equation*}
R_{ab}=\frac{R}{2}p_{ab}.
\end{equation*}
Consequently due to the symmetry between the Ricci tensor and the energy
momentum tensor for a string fluid:%
\begin{equation*}
T_{ab}=\frac{R}{2}(u_{a}u_{b}-n_{a}n_{b})
\end{equation*}
which is in agreement with the quoted result.

\begin{proposition}
\label{Prop SCKV}Let $\xi ^{a}=\xi n^{a}$ be a proper special CKV in an EMSF
space-time and let the total energy of the magnetofluid $\mu +\frac{1}{2}%
\lambda H^{2}+\Lambda \neq 0.$ Then for $\overset{\ast }{H}\neq 0$ we have
the following

(a) The Ricci tensor satisfies the property\footnote{%
It is not necessarily an Einstein space!}:%
\begin{equation*}
R_{ab}=\frac{R}{2}p_{ab}
\end{equation*}

(b) The quantity $R/H$ \ is constant along the magnetic field lines and
along the fluid lines and

(c) The following equations hold:%
\begin{eqnarray*}
\overset{\ast }{\mu }-\left( \mu -\frac{1}{2}\lambda H^{2}+\Lambda \right)
(\ln H)^{\ast } &=&0 \\
\overset{.}{\mu }-(\mu +p_{\perp })(\ln H)^{\cdot } &=&0
\end{eqnarray*}

together with equations (\ref{KC.46}) - (\ref{KC49.b}).
\end{proposition}

Proof

The first part (a) has been shown above.

Concerning (b) we note that (\ref{KC.41}) \ can be written:%
\begin{eqnarray}
\overset{\ast }{\mu }-\left( \mu -\frac{1}{2}\lambda H^{2}+\Lambda \right)
(\ln H)^{\ast } &=&0\implies  \label{KC.54} \\
\overset{\ast }{\mu }-\left( \frac{R}{2}-\lambda H^{2}\right) (\ln H)^{\ast
} &=&0\implies  \notag \\
\overset{\ast }{\mu }+\lambda HH^{\ast }-\frac{R}{2}(\ln H)^{\ast }
&=&0\implies  \notag \\
(\mu +\frac{1}{2}\lambda H^{2}+\Lambda )^{\ast }-\frac{R}{2}(\ln H)^{\ast }
&=&0\implies  \notag \\
(\ln R)^{\ast }-(\ln H)^{\ast } &=&0\implies  \label{KC.53} \\
\left( \ln \frac{R}{H}\right) ^{\ast } &=&0  \notag
\end{eqnarray}

from which follows that the quantity $R/H$ is constant along the magnetic
field lines.

Working in exactly the same way we show that $(\ln (R/H))^{\cdot }=0$ \ from
which follows \ that the quantity $R/H$ is constant along the fluid flow
line.

Concerning the case of KVs we have the following result.

\begin{proposition}
\label{Spacelike KV}An EMSF spacetime admits a KV of the form $\xi ^{a}=\xi
n^{a}$ where $n^{a}=H^{a}/H$ \ iff

(a) $\overset{\ast }{H}=\overset{\ast }{\xi }=0$

(b) The following equations hold%
\begin{eqnarray}
\overset{\ast }{\mu } &=&\overset{\ast }{p_{\perp }}=0  \label{KC.65a} \\
\overset{.}{\mu }-(\mu +p_{\perp })(\ln H)^{\cdot } &=&0  \label{KC.61} \\
\mathcal{S}_{ab} &=&0,\mathcal{R}^{a}=-\frac{j^{b}H_{b}}{2H^{3}}H^{a}
\label{KC.62} \\
\overset{\ast }{n}^{a} &=&-(\ln \xi )^{,a}  \label{KC.63} \\
\overset{.}{n}^{a} &=&p_{b}^{a}\sigma _{c}^{b}n^{c}  \label{KC.64} \\
\sigma _{ab}n^{a}n^{b}-\frac{2}{3}\theta &=&\left( \ln H\right) ^{\cdot }
\label{KC.65} \\
N_{a} &=&0,\text{ \ }e=2\omega ^{a}H_{a}.\text{ }
\end{eqnarray}
\end{proposition}

We conclude that when an EMSF admits the KV $\xi ^{a}=\xi n^{a}$ the
following results hold:

i) Because $n^{a}=\omega ^{a}/\omega =H^{a}/H$ ($\omega \neq 0)$ the string
consists of the 2-dimensional timelike surface spanned by $u^{a}$ and the
vorticity $\omega ^{a}$ (Nambu geometric string) or $\omega ^{a}=0$.

ii) From equations (\ref{KC.28a}), (\ref{KC.28b}) it follows:%
\begin{equation}
L_{\xi }p_{ab}=0,\text{ \ }L_{\xi }h_{ab}=0  \label{KC.55}
\end{equation}%
that is, $\xi ^{a}$ is also a KV\ of the metric $h_{ab}$ \ of the 3-space
normal to $u^{a},$ and a KV of the screen space metric $p_{ab}.$

iii) From Proposition \ref{Theorem 1} we have that the Killing symmetry is
inherited by the geometric and the dynamic variables, that is:%
\begin{equation}
L_{\xi }u_{a}=L_{\xi }n_{a}=0,\text{ \ }L_{\xi }\sigma _{ab}=0,L_{\xi
}\omega _{ab}=0,L_{\xi }\theta =0,L_{\xi }\overset{.}{u}_{a}=0.
\label{KC.56}
\end{equation}

iv) If $\omega ^{a}\neq 0$ then $u^{a},n^{a}=\omega ^{a}/\omega $ must
commute.

Obviously these restrictions are severe and allow only few special choices
for the string fluids in given spacetimes.

\subsection{Application: EMSF in Bianchi I\ spacetime}

\label{section7}

The Bianchi I spacetime with metric

\begin{equation}
ds^{2}=-dt^{2}+A_{1}^{2}(t)dx^{2}+A_{2}^{2}(t)dy^{2}+A_{3}^{2}(t)dz^{2}.
\label{KC.57}
\end{equation}%
has been a platform for studying anisotropy and more specifically string
fluids and electromagnetic fields. For example Letelier \cite{Letelier 1980}
studied string dust in Bianchi I spacetime whereas the electromagnetic field
in the relativistic RMHD has been studied (among many others) in \cite%
{Jacobs Bianchi 1969}. \ Following this line of research we shall use the
results obtained in the last section to compute all possible Bianchi I
spacetimes (if any), which carry a magnetic field satisfying the RMHD\
approximation and admit a spacelike CKV or a spacelike KV.

In order to get comparable results with the literature we consider the
comoving observers $\ u^{a}=(1,0,0,0).$ This choice has a double effect. On
the one hand gives that the vorticity $\omega ^{a}=0$, \ therefore Maxwell
equation $e=2\omega ^{a}H_{a}$ \ implies that the charge density $e=0.$ This
excludes all analytical solutions found in \cite{Jacobs Bianchi 1969}.
Secondly the geometric condition $N^{a}=0$ restricts heavily the possible
symmetry vectors $\xi ^{a}=\xi n^{a}.$ All the CKVs of the Bianchi I
spacetime have been found in \cite{tsbianchi}.

We have checked that for this choice of $u^{a}$ none of these vectors
satisfies the condition $N^{a}=0.$ Therefore the only remaining choice is
the KVs so that the system of equations we have to solve are equations (\ref%
{KC.65a}) to (\ref{KC.66}).

Consider now that $\xi ^{a}=\xi (t)n^{a}$ \ where $n^{a}=\partial
_{z}=(0,0,0,1/A_{3}(t)).$ Equation (\ref{KC.65a}) implies that $\overset{%
\ast }{\rho },~\overset{\ast }{p_{\perp }}~$ are zero hence $\rho
(t),p_{\perp }(t)$. We prove easily that equation (\ref{KC.63}) gives $\xi
=1 $ therefore the KV is the $\ \partial _{z}.$ Equation (\ref{KC.64}) is
satisfied identically, while equation (\ref{KC.65}) gives $%
H(t)=(A_{1}(t)A_{2}(t))^{-1}$.~Therefore the magnetic field is given by%
\begin{equation*}
H^{a}=(A_{1}(t)A_{2}(t))^{-1}\partial _{z}.
\end{equation*}

It remains equation (\ref{KC.61}) \ which is written as%
\begin{equation}
\overset{.}{\rho }+(\rho +p_{\perp })\ln \left[ A_{1}(t)A_{2}(t)\right]
^{\cdot }=0.  \label{KC.69}
\end{equation}

For each equation of state we determine \ a Bianchi I\ space time which
admits a string fluid. For example, let us consider the eqn of state $%
p_{\perp }=\rho \neq 0.$ Then from (\ref{EqnSS.15a}) we have that the EMSF\
has the enrgy momentum tensor%
\begin{eqnarray}
T_{ab} &=&\left( \rho +\frac{1}{2}\lambda H^{2}\right) u_{a}u_{b}-\left(
\rho +\frac{1}{2}\lambda H^{2}\right) n_{a}n_{b}+\underbrace{\left( \rho +%
\frac{1}{2}\lambda H^{2}\right) }_{\equiv q}p_{ab}  \notag \\
&=&\left( \rho +\frac{1}{2}\lambda H^{2}\right)
(u_{a}u_{b}-n_{a}n_{b}+p_{ab})  \label{KC.70}
\end{eqnarray}%
Eqn (\ref{KC.69}) gives 
\begin{equation}
\rho =\frac{c}{A_{1}(t)A_{2}(t)}  \label{KC.71}
\end{equation}%
therefore in the Bianchi I\ spacetime we know the string fluid as well as
the magnetic field.

\section{Conclusions}

\label{section8}

We have applied the 1+1+2 decomposition to the case of the EMSF in the RMHD\
approximation. We have shown that a geometric assumption in the form of a
symmetry effects both the kinematics and the dynamics of the resulting EMSF.
We have approached the problem in two steps a. In full generality
independently of a particular symmetry and b. In the case of a CKV which is
of the form $\xi ^{a}=\xi u^{a}$ and of the form $\xi ^{a}=\xi n^{a}$ with $%
n^{a}=H^{a}/H$ where $u^{a}$ is the four velocity of the fluid and $H^{a}$
is the magnetic field. We applied the results of the $\xi ^{a}=\xi u^{a}$
case in the FRW spacetime and the results of the case $\xi ^{a}=\xi n^{a}$
in the Bianchi I\ spacetime where. In the latter case we found new solutions
for the gravitational field.

It is apparent that the results stated in this work due to their generality
can be used in many different situations involving the electromagnetic field
and various types of symmetries. However one may ask if all the initial
conditions are viable for the solutions which follow from the existence of
symmetries. In particular, the existence of a symmetry in a solution is a
strong argument which its violations leads to another kind of solution. On
the other hand, from the theory of similarity solutions of differential
equations \cite{in1,in2} we know that for a given differential equation a
similarity solution satisfy the initial value problem/boundary conditions
iff the later are also invariant under the action of the symmetries which
provide the similarity transformations. That property can be applied in
order to define initial conditions where a nonsymmetric solution can be
related with a symmetric one. For instance, to relate the inner and outer
solutions in a compact body.

\subsection*{\textbf{Acknowledgement}}

The authors thank the anonymous referees for their comments and suggestions
which helped to improve the quality and the presentation of this work. AP
acknowledges the financial support of FONDECYT grant no. 3160121 and thanks
the University of Athens for the hospitality provided.


\begin{thebibliography}{99}
\bibitem{mhd0} E. Gourgoulhon, (2006) "An introduction to relativistic
hydrodynamics" EAS Publ. Ser. 21, 43

\bibitem{mhd1} S. Grozdanov, D. M. Hofman and N. Iqbal (2017),
\textquotedblleft Generalized global symmetries and dissipative
magnetohydrodynamics \textquotedblright Phys. Rev. D 95, 096003

\bibitem{mhd2} J. Hernandez and P. Kovtum (2017), \textquotedblleft
Relativistic Magnetohydrodynamics \textquotedblright JHEP 1705 (2017) 001

\bibitem{mhd3} J. Armas and A. Jain (2018), \textquotedblleft
Magnetohydrodynamics as superfluidity \textquotedblright arXiv:1808.01939
[hep-th]

\bibitem{let1979} Letelier P (1979), Clouds of Strings in general
Relativity, Phys.\ Rev. D 20, 1294

\bibitem{ss001} L.L. Smally and J.P. Krisch (1996), String fluid dynamics,
Class. Quantum Grav. 13, L19

\bibitem{Ray  (1978)} Ray D (1978) \textquotedblleft Some solutions for
relativistic vortices interacting through a scalar field\textquotedblright\
Phys Rev D, \textbf{18}, 3879 - 3880

\bibitem{Lund F Regge T (1976)} Lund F and Regge T (1976), \textquotedblleft
Unified approach to strings and vortices with soliton solutions" Phys Rev D, 
\textbf{14}, 1524 - 1535

\bibitem{Letelier 1980} Letelier P (1980), \textquotedblleft Anisotropic
fluids with two perfect fluid components\textquotedblright\ Phys Rev D 
\textbf{22}, 807 - 813

\bibitem{Letelier 1981} Letelier P (1981),
\textquotedblleft\textquotedblright\ Nuovo Cimento \textbf{B63}, 519

\bibitem{Letelier 1983} Letelier P (1983), \textquotedblleft String
Cosmologies\textquotedblright\ Phys Rev D \textbf{28}, 2414 - 2419

\bibitem{MHD_1} S. Grozdanov, D. M. Hofman and N. Iqbal (2017),
\textquotedblleft Generalized global symmetries and dissipative
magnetohydrodynamics \textquotedblright Phys. Rev. D 95, 096003

\bibitem{MHD_2} J. Hernandez and P. Kovtum (2017), \textquotedblleft
Relativistic Magnetohydrodynamics \textquotedblright JHEP 1705 (2017) 001

\bibitem{MHD_3} J. Armas and A. Jain (2018), \textquotedblleft
Magnetohydrodynamics as superfluidity \textquotedblright arXiv:1808.01939
[hep-th]

\bibitem{Yavuz Yilmaz (1997)} I Yavuz and I Yilmaz "Inheriting Conformal and
Special Conformal Killing Vectors in String Cosmology" (1997) Gen Rel Grav 
\textbf{9}, 1295 - 1307

\bibitem{Yilmaz 2001} Yilmaz I (2001), ``Timelike and Spacelike Ricci
Collineation Vectors in String Cosmology'' Inter Jour Modern Physics \textbf{%
10}, 681 - 690

\bibitem{Baysal Yilmaz 2002} Baysal H Yilmaz I (2002), ``Spacelike Ricci
Inheritance vectors in a model of string cloud and string fluid stress
tensor'' Class Quantum Grav {\textbf{19}, 6435 - 6443 }

\bibitem{Baysal Camci et all 2002} Baysal H, Camci U, Tarhan I and Yilmaz I
(2002), \textquotedblleft Conformal Collineations in String
Cosmology\textquotedblright\ Inter Jour Modern Physics \textbf{11}, 463 - 469

\bibitem{U Camci 2002} Camci U (2002), \textquotedblleft Conformal
Collineations and Ricci Inheritance symmetry in String Cloud and String
Fluids\textquotedblright\ Inter Jour Modern Physics \textbf{11}, 353 - 366

\bibitem{Sharif M Sheikh U 2005} Sharif M Sheikh U (2005), \textquotedblleft
Timelike and Spacelike Matter Inheritance Vectors in Specific Forms of
Energy-Momentum Tensor\textquotedblright\ Inter Jour Modern Physics (to
appear), gr-qc/0504101

\bibitem{Mason Tsamparlis  1985} Mason D P and Tsamparlis M 1985,
\textquotedblleft Spacelike Conformal Killing Vectors and Spacelike
Congruences\textquotedblright\ J. Math. Phys. \textbf{26}, 2881 - 2901

\bibitem{Maartens Mason Tsamparlis  1985} Maartens R, Mason D P and
Tsamparlis M 1986, ``Kinematic and Dynamic Properties of Conformal Killing
Vectors in Anisotropic Fluids'' J. Math. Phys. \textbf{27}, 2987 - 2994

\bibitem{Saridakis Tsamparlis 1991} Saridakis E and Tsamparlis M 1991,
\textquotedblleft Symmetry Inheritance of Conformal Killing
Vectors\textquotedblright\ J. Math. Phys. \textbf{32}, 1541 - 1551

\bibitem{Noris Green P Davis 1977} Noris L K, Green P. and Davis W R (1977),
J. Math. Phys. \textbf{18}, 1305

\bibitem{Tsamparlis 1992} Tsamparlis M 1992, \textquotedblleft
Geometrization of a General Collineation\textquotedblright\ J. Math. Phys. 
\textbf{33}, 1472 - 1479

\bibitem{Coley Tupper (1989)} Coley A.A and Tupper B. O. (1989) J. Math.
Phys. \textbf{30}, 2616

\bibitem{Tsamparlis Mason 1990} Tsamparlis M and Mason D P 1990,
\textquotedblleft Ricci Collineation Vectors in Fluid
Space-times\textquotedblright\ J. Math. Phys. \textbf{31}, 1707 - 1722

\bibitem{Jacobs Bianchi 1969} Jacobs K (1969), \textquotedblleft Cosmologies
of Bianchi Type I\ with a Uniform Magnetic Field\textquotedblright\ Astro.
Journal \textbf{155}, 379 - 391

\bibitem{Tsamparlis M GRG 2006} Tsamparlis M (2006), \textquotedblleft
General symmetries of a string fluid spacetime\textquotedblright\ Gen Rel
Grav (2006)

\bibitem{dunn} K.A. Dunn and B.O.J Tupper, Astroph. J. 235, 307 (1980)

\bibitem{plasma} V.N. Duarte and R.A. Clement, J. Phys. Conf. Ser. 511,
012015 (2014)

\bibitem{ellis2} G.F.R. Ellis and H. van Elst and, Cosmological models, Carg%
\`{e}se Lectures (1998)

\bibitem{tsbianchi} M. Tsamparlis, A. Paliathanasis and L. Karpathopoulos,
Gen. Relativ. Grav. 47, 15 (2015)

\bibitem{SR} M. Tsamparlis, \textquotedblleft Special Relativity: An
Introduction with 200 Problems and Solutions \textquotedblright , Springer,
2010

\bibitem{in1} P.E.\ Hydo, Symmetry analysis of initial-value problems, J.
Math. Anal. Appl. \ 309, 103 (2005)

\bibitem{in2} R. Cherniha and S. Kovalenko, Lie symmetries of nonlinear
boundary value problems, CNSNS 17, 71 (2012)
\end{thebibliography}
\end{document}